\documentclass[aps,pre,twocolumn,superscriptaddress]{revtex4-1}

\usepackage{latexsym}
\usepackage{amsmath, amsthm, amssymb}
\usepackage{mathrsfs}
\usepackage{epsfig}
\usepackage{graphicx}
\usepackage{dcolumn}
\usepackage{expl3}
\usepackage{float}

\usepackage{tikz}
\usetikzlibrary{trees}

\usepackage{natbib}
\usepackage{hyperref}
\usepackage{placeins}

\allowdisplaybreaks
\begin{document}
\title[Modeling dynamical transitions in somitogenesis]{Morphogen-regulated contact-mediated signaling between cells can drive the transitions underlying body segmentation in vertebrates}
\author{Chandrashekar Kuyyamudi}
\affiliation{The Institute of Mathematical Sciences, CIT Campus, Taramani, Chennai 600113, India}
\affiliation{Homi Bhabha National Institute, Anushaktinagar, Mumbai 400 094, India}

\author{Shakti~N.~Menon}
\affiliation{The Institute of Mathematical Sciences, CIT Campus, Taramani, Chennai 600113, India}

\author{Sitabhra~Sinha}
\affiliation{The Institute of Mathematical Sciences, CIT Campus, Taramani, Chennai 600113, India}
\affiliation{Homi Bhabha National Institute, Anushaktinagar, Mumbai 400 094, India}


\begin{abstract}
We propose a unified mechanism that reproduces
the sequence of dynamical transitions observed during somitogenesis,
the process of body segmentation during embryonic
development, that is invariant across all vertebrate species. This is achieved by
combining inter-cellular interactions mediated via receptor-ligand coupling
with global spatial heterogeneity introduced
through a morphogen gradient known to occur along the
anteroposterior axis. Our model reproduces synchronized oscillations in the
gene expression in cells at the anterior
of the presomitic mesoderm (PSM) as it grows by adding new cells at its posterior,
followed by traveling waves and subsequent arrest of activity, with the eventual
appearance of somite-like patterns.
This
framework integrates a boundary-organized pattern formation
mechanism, which uses positional information provided by a
morphogen gradient, with the coupling-mediated self-organized
emergence of collective dynamics, to explain the processes that
lead to segmentation.
\end{abstract}
\maketitle

\section{Introduction}
The process of development in biological organisms crucially involves the self-organized emergence
of spatial patterns~\cite{Koch1994}. One of the most ubiquitous of such patterns is manifest during
somitogenesis, i.e., the formation of
somites, which are the modular building blocks of all
vertebrate bodies~\cite{dequeant2008,gomez2008,vonk2008}.
Somites compose bilaterally symmetric segments that are formed in the
paraxial, or presomitic, mesoderm (PSM) of developing embryos as
the body axis itself elongates~\cite{Gilbert2013}.
Analogous processes have been implicated in the body segmentation of some
invertebrates~\cite{stollewerk2003,sarrazin2012}.
Although there is great variability across species in terms of the
number of somites, the mean
size of a somite and the duration over which they are
formed, nonetheless a conserved set of features characterizing
somitogenesis is seen across these species\hspace{0.1cm}\cite{pourquie2001}.
A general conceptual model
for explaining these core features is provided by the Clock and Wavefront (CW)
framework proposed by Cooke and Zeeman in 1976~\cite{cooke1976},
that allows the translation of a temporal sequence into a spatial pattern~\cite{pourquie2003}.
They assumed the PSM to comprise
cellular oscillators (clocks) which are each arrested at their instantaneous
state of activity upon
encountering a wavefront that moves from the
anterior to the posterior of the PSM~\cite{baker2006,santillan2008,nagahara2009,hester2011,ares2012,murray2013,jorg2014,wiedermann2015}.

In order to construct an explicit mechanism embodying the CW framework,
we need to disaggregate its components that operate at different
length scales, namely, (i) the cellular scale at which oscillations occur,
(ii) the inter-cellular scale at which contact-mediated
signaling takes place, and (iii) the scale of the PSM across
which morphogen gradients form and act as the environment
that could modulate the inter-cellular interactions.
This resonates with the proposal of Oates~\cite{Oates2012} to view the
CW framework as a three-tier process.
In the bottom tier, we observe oscillations at the level of a single cell
in the PSM, arising from the periodic expression of clock genes~\cite{palmeirim1997,dale2000,saga2001,maroto2003,masamizu2006,riedel2007,schroter2012,webb2016}.
The middle tier describes the mechanism by which the cellular oscillators
coordinate their activity with that of their neighbors. This occurs through
juxtacrine signaling brought about by interactions between
Notch receptors and Delta
ligands~\cite{jiang1998,ferjentsik2009,hubaud2014,conlon1995,pourquie1999,yun2000,lai2004,maroto2003,huppert2005,mara2007,kageyama2007,sprinzak2010,sprinzak2011}.
Indeed, several earlier models have
explored the role of Notch-Delta coupling in bringing about robust
synchronization between the oscillators~\cite{lewis2003,giudicelli2004,horikawa2006,Giudicelli2007,tiedemann2012,tiedemann2014}.
Finally, processes that bring about the slowing down (and eventual
termination) of the oscillations~\cite{yun2000,mcgrew1998,palmeirim1997},
and the subsequent differentiation of the cells into rostral
and caudal halves of the somites~\cite{Oginuma2010}, constitute the top tier.
\begin{figure}
\centering
\includegraphics[width=0.99\columnwidth]{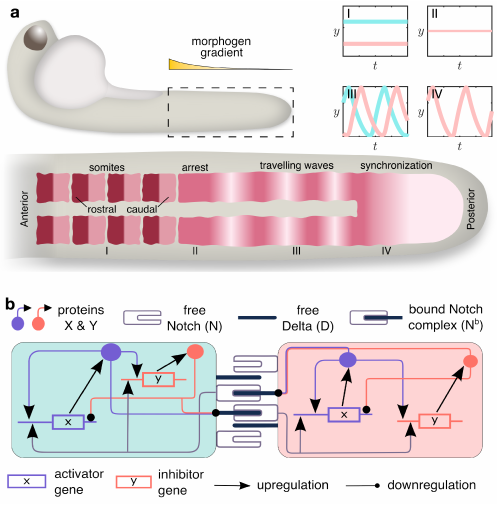}
\caption{\textbf{The key dynamical features of somitogenesis (top)
that are reproduced in our mathematical model (bottom).}
(a) Schematic diagram depicting the zebrafish
presomitic mesoderm (PSM) and the dynamical states exhibited by the
cells over the course of somitogenesis. The box outlined with broken lines
represents the
posterior region which exhibits a spatial gradient in
morphogen concentration,
represented by the exponentially varying profile
shown above the box. The dynamics of the cells at the tail (region IV)
are characterized by exact synchronization of the periodic variation
in gene expression. As the cells move towards
the anterior, they first exhibit travelling waves of gene expression (III),
followed by arrest of the oscillations (II). Eventually the cells
differentiate (I) into alternating bands corresponding to rostral
and caudal halves of the mature somites.
The figures in the insets at the right
display typical time series in each of the regimes I-IV of the inhibitor
gene expression for two neighboring cells coupled to each other.
(b) Schematic diagram
describing the interaction between two cells via
Notch-Delta coupling. In general, each cell has Notch receptors, as well
as, Delta ligands that bind to them. The \textit{cis} form of the binding leads
to the loss of receptors and ligands without resulting in any downstream signaling, whereas \textit{trans} Notch-Delta
binding gives rise to cleavage of the Notch Intra-Cellular Domain (NICD).
The latter acts as a transcription factor (TF) for the downstream
activator ($x$) and inhibitor ($y$) genes which are the essential
constituents of each cellular oscillator. The proteins $X$ and $Y$ resulting from the expression
of these genes, in turn, downregulate the production of Delta ligands (the repression
being indicated by arrows with circular heads).
}
\label{fig:fig1}
\end{figure}
\begin{figure}
\includegraphics[width=0.95\columnwidth]{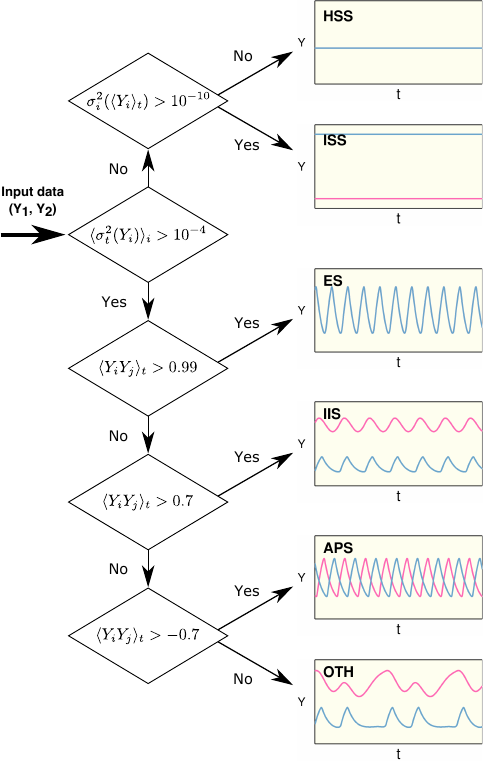}
\caption{{\bf Flow chart illustrating the algorithm used to classify the
collective dynamical patterns obtained for a system of two coupled oscillators.}
To distinguish between oscillating
and non-oscillating patterns, we use $\langle \sigma^{2}_{t} (Y_{i})\rangle_{i}$
which is the mean temporal variance of the time series of $Y$, the concentration
of the protein product of gene $y$, calculated
over the two oscillators. To determine whether the steady states that the
oscillators have reached are the same (corresponding to
HSS) or different (corresponding to ISS), we compute the
variance of the mean values for the two time series,
$\sigma^{2}_{i} (\langle Y_{i}\rangle_{t})$. To distinguish between the
oscillating patterns, we use the equal time linear correlation between the two
time series, $\langle Y_{i} Y_{j}\rangle_{t}$.
The classification is robust with respect to small changes in the values of
the thresholds, which are displayed in the figure.}
\label{fig:flowchart}
\end{figure}

In this paper we propose a model that
integrates these different length scales by investigating genetic oscillators
interacting via Notch-Delta coupling whose strength is modulated in a
position-dependent manner due to a morphogen
concentration gradient along the anteroposterior (AP) axis of the PSM.
This provides an unified framework for explaining the dynamical transitions
observed during somitogenesis.
As the PSM expands along the AP axis through the addition of new cells at
the tail~\cite{Gilbert2013}, it is reasonable to restrict our attention
to the process of somitogenesis taking place along a one-dimensional array
of coupled cells in the PSM.
From the perspective of our modeling which explicitly
investigates the role of morphogen gradient in coordinating somite formation,
the array of cells is considered to be aligned along the AP axis, as
the various morphogens that are expressed along the PSM are known
to form concentration gradients along this axis~\cite{gibb2010}.
These primarily include molecules belonging to the
FGF (Fibroblast Growth Factor)~\cite{dubrulle2001,dubrulle2004},
RA (Retinoic Acid)~\cite{dubrulle2004development,vermot2005} and
Wnt (Wingless/integrated) families~\cite{aulehla2004,aulehla2008,bajard2014}.
Even though the role of gradients on the overall dynamics has been
explored~\cite{goldbeter2007,goldbeter2008,mazzitello2008,jorg2016}, there is
to date no consensus as to the explicit mechanism through which they contribute
to somite formation.
Our model demonstrates that if the
morphogen gradient is considered to regulate the impact of the Notch intracellular domain (NICD)
on the expression
of the clock genes, it can lead to qualitatively different kinds of
dynamics along the PSM in a threshold-dependent
manner [Fig.~\ref{fig:fig1}~(a)].
The dynamical evolution of our model culminates with the emergence of somites,
each comprising two cells, that resemble the empirically observed segments that
occur towards the tail end of the mesoderm~\cite{Youn1980,Tlili2019}.
Unlike the conventional boundary-organized pattern
formation paradigm~\cite{Lander2011}, here the morphogen gradient
does not determine the cell fate so much as affect
the interaction between neighboring cells that lead to
dynamical transitions similar to those observed in somitogenesis.
This is in contrast to previous work where spatial heterogeneity introduced by the
morphogen gradient is incorporated through variations in the autonomous oscillatory behavior of individual cells~\cite{murray2011,Tomka2018}.
Also, while it has been suggested that the mechanical deformations that the tissue undergoes during development could play a role in somitogenesis~\cite{Adhyapok2021,Narayanan2021},
our work shows that its broad features can be explained exclusively by the interactions
between cells and the morphogen signal.
While earlier studies have reproduced the different spatio-temporal patterns that
arise over the course of somite formation~\cite{Meinhardt1982,Francois2007,cotterell2015}, the significance of the model presented here centers around demonstrating that
morphogen gradients may play a crucial role in regulating inter-cellular Notch-Delta mediated
interactions, whose role in somitogenesis has been established experimentally~\cite{Wahi2016,liao2017}.
Thus, our results help address an open question as to how
morphogen gradients influence the collective dynamics
of the cellular oscillators by differentially modulating the inter-cellular coupling both in space and time.
\begin{figure}
\centering
\includegraphics[width=0.99\columnwidth]{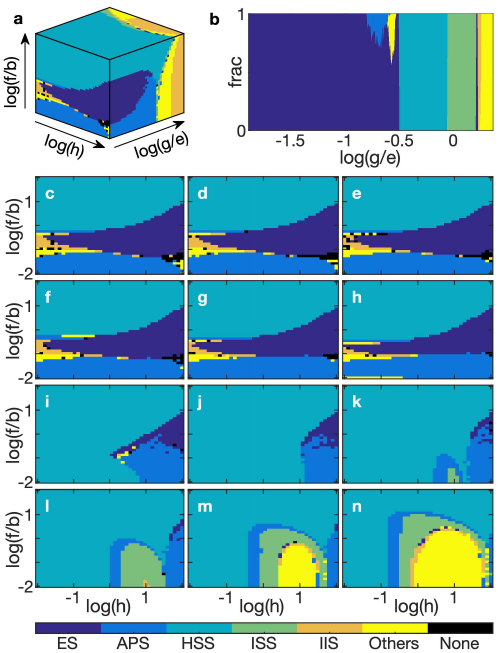}
\caption{\textbf{Transitions between patterns representing different
stages in somitogenesis seen in
the collective dynamics of a pair of cells on varying the
parameters governing the strength of Notch-Delta coupling in our model.}
(a) Schematic diagram of the three-dimensional space spanned by the
coupling parameters ($f, g, h$), scaled by the relevant kinetic parameters of
the individual oscillators. Note the logarithmic scale used for the
ranges of the parameters.
(b) The variation with $g$ of the relative frequency of occurrence of
patterns belonging to each
of six distinct categories, viz., Exact Synchronization (ES), Anti-Phase
Synchronization (APS), Homogeneous Steady States (HSS),
Inhomogeneous Steady States (ISS), Inhomogeneous In-phase Synchronization
(IIS) and Other dynamical patterns (Others) [``None'' refers to those regions of parameter space in which
no single pattern dominates].
The values of the parameters $f/b$ and $h$ are fixed at $0.25$ and $4$,
respectively.
(c-n) The most commonly occurring dynamical patterns (i.e., obtained for
$>50\%$ of all initial conditions used) that are seen for different
values of $f/b$ and $h$ (varying over four order of magnitude) for
$12$ equally spaced values of $\log (g/e)$ between $-1.89$ (panel c) and
$0.36$ (panel n).
While HSS is the most common pattern seen over this range of parameter
values, focusing on how the occurrence frequency of ES varies with $f,g,h$
indicates where a transition from synchronization to time-invariant behavior
may be achieved.
In all cases, initial values of the dynamical variables for each cell are independently
and identically distributed uniformly over the unit interval.
Unless mentioned otherwise, the following parameter values have been used for
all model simulations: $a = 16.0$, $b = 200.0$, $c = 20.0$, $e = 10.0$, $\beta^N = 5.0$, $\beta^D = 100.0$, $\gamma =1.0$, $k^{tr} = 1.0$, $k^{cis} = 1.0$ and $\mu = 1.0$.
}
\label{fig:fig2}
\end{figure}

\section{Methods}
\subsection{Modeling genetic oscillators interacting via Notch-Delta coupling}
Several experiments have established that the cells in the presomitic
mesoderm (PSM) have ``clock'' genes whose expression levels
oscillate~\cite{palmeirim1997,dale2000,saga2001,maroto2003,masamizu2006,riedel2007,webb2016}.
In our model, we consider a generic two component genetic oscillator
comprising an activator gene ($x$), which upregulates its own expression,
as well as that of an inhibitor gene ($y$), which suppresses the expression
of the activator gene~\cite{Guantes2006}.
The dynamics of this two-component
oscillator can be expressed in terms of a pair of rate equations
describing the change in concentrations of the protein products $X$ and
$Y$ of genes $x$ and $y$, respectively (which, in the case of zebrafish, can be identified with the \textit{her1} and \textit{her7}
genes~\cite{Oates2012,lewis2003}).
The model parameters are chosen
such that $X$ and $Y$ exhibit limit cycle oscillations (see Supplementary Information, Fig.~S1).

As the communication between cells in the PSM is crucial in mediating
their collective behavior during somitogenesis,
we couple the dynamics of the clock genes of neighboring cells.
Experiments have established the role of the Notch-Delta juxtacrine
signaling pathway in mediating the interaction between cells that are
in physical contact with each other~\cite{jiang1998,ferjentsik2009,hubaud2014,conlon1995,pourquie1999,yun2000,lai2004,maroto2003,huppert2005,mara2007,kageyama2007}.
Such a receptor-ligand based mechanism is crucial for allowing communication between cells
during processes such as somitogenesis, as
gene products (proteins or mRNA) are too large to be transported across the cellular
membrane, thereby preventing direct interaction between cells via diffusion~\cite{Goodenough2009}.
In general, each cell has both Notch receptors as well as Delta ligands
on their surface. A Notch receptor on cell $i$ which is bound to a
Delta ligand belonging to a neighboring cell $j$ (i.e., \textit{trans} binding)
leads to the cleavage of NICD that will
act as transcription factor for downstream genes in cell $i$~\cite{Takke1999,Oates2002}.
In our model, following Refs.\cite{lewis2003,horikawa2006}, NICD upregulates the expression of both the
clock genes, while the gene products $X, Y$ suppress the production of
Delta ligands by the cell. We describe
the Notch-Delta signaling mechanism through the coupled dynamics of (i) the
free (unbound) Notch receptor concentration ($N$), (ii) the free Delta
ligand ($D$) and (iii) the NICD which is released as a result of \textit{trans}
binding ($N^b$) [see the schematic diagram shown in the Supplementary Information, Fig.~S3].
Thus, the dynamics of a cell $i$ coupled to its neighbors through Notch-Delta
signaling is described by the following set of equations:
\begin{eqnarray}
\label{eq:fullmodel1}
\frac{dX_i}{dt} &=&  \frac{a + b X_{i}^2 + f N_{i}^{b}}{1 + X_{i}^{2} + Y_{i}^{2} + N_{i}^{b}} - cX_{i}\,, \\
\label{eq:fullmodel2}
\frac{dY_i}{dt} &=&  \frac{e X_{i}^2 + g_i (t) N_{i}^{b}}{1 + X_{i}^{2} + N_{i}^{b}} - Y_{i}\,, \\
\label{eq:fullmodel3}
\frac{dN_i}{dt} &=&  \beta^{N} - \gamma N_i - k^{\rm cis}D_iN_i - k^{\rm tr}D^{\rm tr} N_i\,, \\
\label{eq:fullmodel4}
\frac{dN_{i}^{b}}{dt}  &=& k^{\rm tr}D^{\rm tr}N_i - \mu N_{i}^{b}\,, \\
\label{eq:fullmodel5}
\frac{dD_i}{dt} &=&  \frac{\beta^{D}}{1 + h\,(X_{i}^{2} + Y_{i}^{2})} - \gamma D_i - k^{\rm cis}D_i N_i - k^{\rm tr}D_i N^{\rm tr}\,.
\end{eqnarray}
Here the terms $D^{tr}$ and $N^{tr}$ are the mean values of $D_{j}$
and $N_{j}$ over all neighboring cells $j$ to which $i$ is coupled
through \textit{trans}-binding.
The functional form chosen for the terms corresponding to binding interactions in Eqs.~(\ref{eq:fullmodel1}),(\ref{eq:fullmodel2}),
and (\ref{eq:fullmodel5}) represent the fact that the transcription factors compete with
each other to bind to the same site in
the regulatory regions of the genes coding for $X$, $Y$ and $D$, respectively.
While the values of the model parameters
can, in general, vary across cells,
we restrict our attention to the spatio-temporal variation of the coupling parameter
$g$ [subscripted with the cell index in
Eqn.~(\ref{eq:fullmodel2})],
which determines the strength of upregulation of $y$ by the NICD ($N^b$).
This allows us to
investigate the role of spatial heterogeneity imposed by the gradient of
morphogen concentration
along the anteroposterior (AP) axis of the PSM.
For the simulations reported here we have considered the case $k^{\rm tr}=k^{\rm cis}=1$.
We note that our results are qualitatively unchanged in the absence of {\it cis}-inhibition
(see Supplementary Information, Fig.~S4).
\subsection{Morphogen gradients in the PSM}
It is known that the morphogens RA, Wnt and FGF are differentially expressed
along the PSM, exhibiting monotonically varying concentration gradients
having peaks at the posterior (for FGF and Wnt) or anterior (for RA)
ends~\cite{gurdon2001,aulehla2008,aulehla2010}.
Experiments on several vertebrate species
have shown that high concentrations
of RA initiate differentiation, while increased levels of Wnt and FGF,
which are known to promote sustained oscillations in the expression
of the clock genes, impede the formation of mature
somites~\cite{dubrulle2001,sawada2001,moreno2004,vermot2005}.
Note that some aspects of the roles played by Wnt and FGF are already
accounted for in the local dynamics of our model, where each cell is capable of autonomous robust
oscillations. As our primary goal is to explicate the mechanisms driving termination of oscillatory activity
followed by cellular differentiation, we focus on the role of RA on the collective dynamics of cells in the PSM.
We incorporate the effect of this morphogen in the spatial variation
of the coupling parameter $g$ which is assumed to exponentially decay
from the anterior to the posterior end of the domain. Such a profile will
naturally arise if the morphogen diffuses from a source located at the
anterior and is degraded at a constant rate across space~\cite{Lander2002,Bergmann2007,Barkai2009}.
The regulation of the Notch-mediated interaction can
come about by the binding of a morphogen molecule to a cell surface receptor leading to
(either directly or indirectly) the expression of molecules that aid NICD or its downstream
effector to bind more strongly to the promoter site of gene $Y$.
This contrasts with earlier studies (e.g., see Ref.~\cite{Hubaud2017}) that have assumed the morphogen to regulate the gene expressing Notch, which would affect
expression of both $X$ and $Y$
(instead of selectively affecting only $Y$ as in the present model).
\subsection{Dynamical evolution of the morphogen gradient}
Our model focuses on the behavior of a contiguous segment of cells
of length $\ell$ in
the PSM with a morphogen source located at its anterior. If the strength
of the source is constant in time, it would
have resulted in the gradient becoming progressively less steep as the PSM
expands. This dilution is countered by the net increase in the strength
of the source through the secretion of morphogen by the newly matured somites
(see Supplementary Information, Fig.~S6).
Thus, as the PSM expands due to addition of cells at the posterior tail,
we can view the segment under consideration as effectively flowing up a
morphogen gradient along the AP axis~\cite{Oates2012}.
We choose a segment of $N$ cells with a spatial extent $\ell$
that is initially located ($t=0$) at the posterior
end of the PSM, i.e., at the lower end of the gradient.
Its evolution is followed for a duration $T$, the
time required for the array of cells to move across the entire spatial extent of the morphogen gradient considered here. Thus, it determines the rate
at which cells move along the gradient, which in turn is related to the rate at which
the PSM expands by cell division at its posterior end.
As mentioned above, the effect of the varying morphogen concentration on
the dynamics of the cells is introduced via the coupling parameter $g$.
Specifically, we assume that the value of $g$ at each site is proportional
to the corresponding morphogen concentration, yielding an exponentially
varying gradient of $g$ across the AP axis:
$g_i(t) = g_{\min} \exp\left(\lambda_g x_i(t)\right)$.
The steepness of the gradient is quantified by $\lambda_g$, which is a
function of $T$, as well as $g_{\max}$ and $g_{\min}$, which are the values
of $g$ at the anterior and posterior ends of the PSM, respectively, viz.,
$\lambda_g = \ln(g_{\max}/g_{\min})/T$.
We assume that the effective flow of the segment of cells along the AP axis
occurs at an uniform rate. This can be taken to be unity without loss
of generality by appropriate choice of time unit. Thus, the instantaneous
position $x_i (t)$ along the gradient of the $i$th cell in the segment is
given by $x_i (t) = t + (\ell/N) (i-1)$, with the initial
condition as $x_1(t = 0) = 0$.

\section{Results}
In our simulations, we have considered the PSM to comprise cells, each
of which exhibits oscillating gene expression. We assume a minimal
model for the genetic oscillator
consisting of two clock genes, one activatory and the other inhibitory,
whose products correspond to fate determining proteins
(see Methods). The oscillations of neighboring cells influence each other
through Notch-Delta inter-cellular coupling [Fig.~\ref{fig:fig1}~(b)].
We have explicitly verified that incorporating delay in the contact mediated
signaling does not alter our results qualitatively (see Supplementary Information, Fig.~S5).
\subsection{Dynamics of a pair of coupled cells}
In the simplest setting, namely, a pair of adjacent cells, which
allows us to investigate the effect of coupling on the collective dynamics,
the system can exhibit a wide
range of spatio-temporal patterns.
These can be classified systematically through the use of quantitative
measures (see Fig.~\ref{fig:flowchart}).
We focus on the patterns that can be immediately interpreted in the
context of somitogenesis: (i) Inhomogeneous Steady States (ISS), (ii)
Homogeneous
Steady States (HSS), (iii) Anti-Phase Synchronization (APS) and (iv)
Exact Synchronization (ES) [shown schematically as insets
of Fig.~\ref{fig:fig1}~(a)]. The range of values of the coupling-related parameters $f$, $g$ and $h$
over which these patterns are observed in a pair of coupled cells are shown
in Fig.~\ref{fig:fig2}. The parameters
$f$ and $g$ govern the strength with which the Notch intra-cellular domain
($N^b$) regulates
the activatory and inhibitory clock genes, respectively, while $h$
is related to the intensity of repression of the Delta ligand ($D$)
by each of the clock genes. As inter-cellular coupling is believed to
be responsible for the synchronized activity of cells in the initial
stage of somitogenesis~\cite{maroto2005}, we note that the dynamical regime corresponding
to ES occurs for low $g$ and intermediate values of $f$, with the
region increasing in size for larger $h$.
\begin{figure}
\centering
\includegraphics[width=0.99\columnwidth]{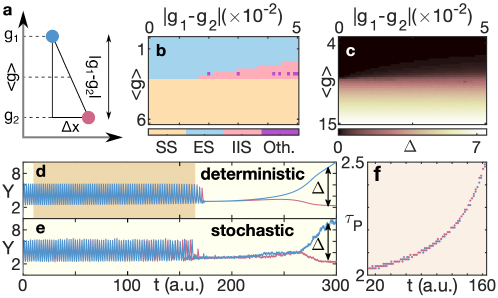}
\caption{\textbf{Incorporating a morphogen gradient by varying the
Notch-Delta coupling parameter $g$ of adjacent oscillators reproduces
the temporal sequence of patterns observed during somitogenesis.}
(a) Schematic representation of the spatial variation of $g$, resulting from
the different concentrations of a morphogen sensed by neighboring cells,
which are separated by a distance $\Delta x$ in space. The
steepness of the gradient is quantified by $|g_{1}-g_{2}|$, the difference in
the values $g_{1,2}$ for the two oscillators, while their location on the
gradient is determined by the mean $\langle g\rangle$.
(b) The most commonly occurring dynamical patterns (i.e., obtained for $>50\%$
of all initial conditions used) on varying $|g_{1}-g_{2}|$ and
$\langle g\rangle$.
The states of the adjacent cells, characterized by the protein concentration $Y$ ,
converge to fixed points $Y_{1,2}$ in the steady state (SS) region.
(c) The difference between the steady state concentrations of the
adjacent cells, $\Delta = |Y_1 - Y_2|$, increases with $\langle g\rangle$
as one effectively moves up the morphogen gradient, while being relatively
unaffected by the steepness $|g_{1}-g_{2}|$.
(d-f) The dynamical consequences of a morphogen gradient with
an exponentially varying concentration profile. (d) As a pair of
coupled cells gradually move upstream of the gradient,
resulting in an increase of $g_{1,2}$ over time $t$,
their collective behavior (represented by $Y$)
converges from initially synchronized oscillations (ES) to
an inhomogeneous steady state (ISS, characterized by
finite values of $\Delta$) at long times.
These transitions are robust with respect to stochastic fluctuations
(as shown in panel e).
The changes occurring in the system during the transition from
oscillations to steady state behavior (shaded region)
can be quantitatively investigated by focusing on how the
period $\tau_p$ of the oscillations
changes over time.
(f) As cells move upstream of the
morphogen gradient (corresponding to a progression from posterior to
anterior regions in the PSM),
the model displays an increase in time period $\tau_p$.
This is consistent
with a key experimental observation, viz., slowing of
oscillations as cells approach the anterior end of the PSM, during
somitogenesis. The coupling parameters are
$f = 50$, $h = 4$, $g_{\max} = 15$ and $g_{\min} = 0.5$, with $\Delta x = 3$ arb. units.
}
\label{fig:fig3}
\end{figure}

\subsection{Morphogen gradient-induced heterogeneity in a pair of coupled cells}
Spatial variation in these coupling parameters across the PSM can arise through
heterogeneity in the underlying morphogen concentrations.
Introducing heterogeneity through the coupling parameter $g$ in the pair of
adjacent cells considered earlier
($g_1$ and $g_2$ being their respective values), we observe that qualitatively
similar spatio-temporal patterns to those observed in Fig.~\ref{fig:fig2}
are obtained. On varying the
mean value of the coupling $\langle g\rangle=(g_{1}+g_{2})/2$, which
effectively represents the location of these cells on the PSM, and the
steepness of the gradient $|g_1-g_2|$ [Fig.~\ref{fig:fig3}~(a)], the
range of $\langle g\rangle$ over which ES is seen (corresponding
to the region proximal to the posterior end of the PSM) does not appear
to change
appreciably on increasing $|g_1-g_2|$ [Fig.~\ref{fig:fig3}~(b)].
Above a critical value of $\langle g\rangle$ which is independent of
the gradient, the activities of the cells are arrested at $Y_1$ and $Y_2$,
respectively, with the gap $\Delta = |Y_1 - Y_2|$ becoming larger
as we move towards the
anterior end, corresponding to increasing
$\langle g\rangle$ [Fig.~\ref{fig:fig3}~(c)].

As explained in the Methods, over the course of development, the PSM expands
through new cells being added to its posterior end, such that the existing
cells progressively encounter increasing values of the morphogen concentration.
Modeling this time-evolution as an effective flow of the segment of
adjacent cells along the gradient in $g$, we observe that
a transient phase of ES is followed by desynchronization and subsequent
attenuation of the oscillations, eventually leading to a separation
of the steady states of the two cells [Fig.~\ref{fig:fig3}~(d); see
also Supplementary Information, Fig.~S7)].
The gap $\Delta$ between the steady states increases with time, giving
rise to a pronounced ISS state.
This duration depends sensitively on
the steepness of the morphogen concentration gradient (see Supplementary Information, Fig.~S8).
The sequence of dynamical transitions seen in the model are robust with
respect to the presence of noise as shown explicitly in Fig.~\ref{fig:fig3}~(e)
by introducing stochastic fluctuations in the dynamical variables (see also Supplementary Information, Fig.~S9).
Immediately preceding the arrest of periodic activity, we observe that the
period $\tau_P$ [Fig.~\ref{fig:fig3}~(f)] of the
oscillations increases with time that is in agreement with
experimental observations of somitogenesis~\cite{delaune2012,shih2015}.
\subsection{Dynamics of a cellular array in a growing PSM in presence of a morphogen gradient}
\begin{figure}
\centering
\includegraphics[width=0.99\columnwidth]{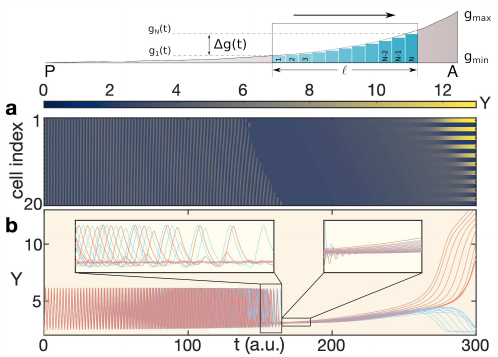}
\caption{\textbf{Collective dynamics of a cellular array responding
to an exponential gradient of morphogen concentration
reproduces the spatio-temporal evolution of PSM activity seen
during somitogenesis.}
The transition from progenitor cells in the posterior (P)
to maturity at the anterior (A) of a segment of length $\ell$ comprising
$N$ cells in the PSM
viewed as a flow upstream (moving window in the schematic on top) over
a period of time $T$
along the exponential profile of the parameter $g$, decaying from
$g_{\max}$ to $g_{\min}$, reflecting the
morphogen gradient.
Different cells in the window sense different morphogen concentrations,
whose values change over time. This is incorporated in terms of
the time-dependent gradient $\Delta g (t)$, with $g_1 (t)$ and $g_N (t)$
being the values of the coupling parameter $g$ at the anterior and
posterior ends of the window at time $t$, respectively.
The resulting change in the activity of a segment comprising $N = 20$
coupled cells, as it moves from P to A, is shown in (a).
The system initially exhibits synchronized oscillations across the segment
but, as a consequence of the gradient, a phase lag develops between adjacent
cells (as seen in the time series in panel b, where odd and even numbered cells are represented
using red and blue curves, respectively.).
This results in a wave-like propagation of the peak
expression from the anterior to the posterior end of the segment in each cycle.
Subsequently, the oscillations reduce in intensity leading to arrest of the oscillations,
a magnified view of the transition
being shown in the left inset of panel (b).
This is followed by a divergence of the dynamical trajectories followed by the different cells, as shown in the right inset of (b).
This gives way to an inhomogeneous steady state (ISS) with adjacent cells attaining different fates characterized by alternating
high and low values of the protein concentration $Y$ [cells with
odd and even indices on the segment are shown using different colors
in panel (b)]. Results shown are for parameter values of $g_{\max} = 15$,
$g_{\min} (= g_0) = 0.5 $ and $T = 300$ a.u.
In all cases, initial values of the dynamical variables for each cell are independently
and identically distributed uniformly over the unit interval.
We note that NICD concentration at each cell exhibits qualitatively similar dynamics (see Supplementary Information, Fig.~S10).
}
\label{fig:fig4}
\end{figure}

Having seen that a pair of contiguous cells can indeed converge to markedly
different steady state values of their clock gene expressions, we now
investigate the generalization to a spatially extended segment of length $\ell$
in the growing
PSM subject to a morphogen gradient [varying from $g_{\min}$ to $g_{\max}$ as
shown schematically above
Fig.~\ref{fig:fig4}~(a)].
As the principal variation of the morphogen concentration occurs along
the AP axis of the PSM, we restrict our focus to a one-dimensional array
of cells aligned along this axis.
As new cells are added to the posterior of the PSM over time,
the relative position of the
segment of cells under consideration shifts along the morphogen gradient
from the posterior to the anterior.
We note that had there been a temporally invariant source of morphogen
at the anterior, the expansion of the PSM would have resulted in a
dilution of the gradient.
However, newly matured somites at the anterior end serve
as additional sources of the morphogen over
time~\cite{del2003,del2004,rhinn2012}, thereby
ensuring that each cell
experiences an exponential increase in the morphogen concentration.
This is reproduced in our model by the segment effectively flowing up the morphogen
gradient (as discussed in Methods).

As shown in Fig.~\ref{fig:fig4}~(a), for a range of values of the
parameters $g_{\max}$, $g_{\min}$ and $\ell$, the cells display a
short-lived ES pattern, which is followed by the development of
a phase lag between adjacent cells [as can be seen in the inset of
Fig.~\ref{fig:fig4}~(b)]. This is analogous to the appearance
of a small phase difference between the pair of oscillators described
earlier, and manifests as a travelling wave
that propagates along the PSM.
We note that such a traveling wave of gene expression has indeed
been experimentally observed to move through the PSM towards the anterior~\cite{Oates2012,masamizu2006,aulehla2008}.
As the cells move further up the gradient, the oscillations subside,
eventually giving way
to a heterogeneous steady state characterized
by adjacent cells having
alternating high and low clock gene expressions. The gap between these high
and low values increases with time to eventually produce a distinctive
pattern that resemble the stripes that arise due to polarization of each
somite into rostral and caudal halves [as seen for large $t$
in both Figs.~\ref{fig:fig4}~(a) and (b)].
In this asymptotic steady state, the separation between the
high and low values for clock gene expression is greater than the amplitude
of the oscillations seen at lower values of $t$.
Thus, we can reproduce the
entire sequence of dynamical transitions observed in the PSM
during somitogenesis
through a model incorporating an array of
oscillators that interact via Notch-Delta signaling while ``moving up'' a
morphogen gradient.
We note that, in different organisms, the size of somites vary across the mesoderm,
with the ones occurring at the posterior end being smaller~\cite{Schroter2008} and resembling the segments consisting of pairs of cells that arise in our model.
We would like to point out that the dynamics resulting from the inter-cellular interactions
can yield ISS states having larger spatial periodicities, which suggests a potential for producing
larger somites (see Supplementary Information, Fig.~S11).
We also note that introducing additional mechanisms such as lateral induction via Jagged receptors~\cite{Hadjivasiliou2016} or diffusive coupling between cells via gap junctions~\cite{Meinhardt1982,cotterell2015} may allow for variation in the wavelength of the periodic pattern.

\section{Discussion}
Somitogenesis is seen across all vertebrates, and recent evidence
implies that mechanisms underlying it could have analogues even
in segmentation of invertebrates, such as
arthropods~\cite{stollewerk2003,clark2019}.
It would appear that there is an invariant set of mechanisms
responsible for this process, that
differ only in terms of the specific identities of the contributing
molecular players across species.
Thus, somitogenesis would in general involve (i) a cellular ``clock'', (ii)
means by which neighboring clocks communicate, and (iii) a spatial gradient
of signaling molecules, which introduces heterogeneity in the interactions
between the clocks.
We have shown here that incorporating these three elements in a model
of a PSM, that grows through
the addition of cells at the posterior,
reproduces the sequence of invariant dynamical transitions seen in
somitogenesis.

While the roles of interacting clocks and that of morphogen gradients
have been investigated individually in earlier studies, we provide here
a framework to understand how these two work in tandem to give rise
to the key features associated with somitogenesis.
In particular, our results shed light on the significance of the
steepness of the morphogen gradient. For instance, we may consider
the consequences of a reduction in the steepness leading to a linear
profile for the morphogen gradient which can arise, for example, when
the degradation rate is negligible.
On replacing the exponential morphogen gradient in our model with a linear one,
we observe a very long-lived transient state before the system converges
to an inhomogeneous steady state. This therefore suggests that
exponential gradients allow relatively
rapid switching between qualitatively distinct dynamical regimes
Hence, by varying the
steepness of the RA gradient experimentally it should be possible to
determine how the time required
for maturation changes as a consequence. This is especially true in the case of
the time interval between cessation
of oscillations and the polarization of the somites.
As inter-cellular coupling is also known to regulate the period of the
segmentation clock~\cite{herrgen2010}, it is possible that introducing
other morphogen gradients, that influence the strength
of the coupling, can explain variations in the rate at which somites form
over time.
The broad features observed here can be reproduced
in two-dimensional cellular arrays with anisotropic inter-cellular coupling (see Supplementary Information, Fig.~S12).
Furthermore, the core assumption of our model, namely that Notch-Delta
coupling plays a crucial role in regulating
somitogenesis in the presence of a morphogen gradient can be
probed in experimental systems where Notch signaling has been arrested.
Future research involving incorporation of additional details in the model
presented here may provide answers to several challenges that explanations
of somitogenesis based on the clock-and-wavefront mechanism have faced~\cite{Stern2015}.

\begin{acknowledgements}
We would like to thank Krishnan Iyer, Jose Negrete Jr, Shubha Tole and Vikas Trivedi
for valuable suggestions. The authors would like to acknowledge discussions during the ICTP/ICTS Winter
Schools on Quantitative Systems Biology (ICTP/smr2879, ICTS/qsb2019/12).
SNM has been supported by the IMSc Complex Systems Project ($12^{\rm th}$
Plan), and the Center of Excellence in Complex Systems and Data Science,
both funded by the Department of Atomic Energy, Government of India.
The simulations and computations required for
this work were supported by High Performance Computing facility
(Nandadevi and Satpura) of The Institute of Mathematical Sciences,
which is partially funded by DST.
\end{acknowledgements}

\clearpage

\onecolumngrid

\setcounter{figure}{0}
\renewcommand\thefigure{S\arabic{figure}}
\renewcommand\thetable{S\arabic{table}}

\vspace{1cm}

\begin{center}
\textbf{\large{SUPPLEMENTARY INFORMATION FOR} \\~\\ Morphogen-regulated contact-mediated signaling between cells can drive the transitions underlying body segmentation in vertebrates}
\end{center}

\section*{List of Supplementary Figures}

\vspace{-0.1cm}
\begin{enumerate}
  \item Fig S1: Schematic diagram of the genetic oscillator model and phase portrait of its dynamics.
  \item Fig S2: Representative patterns of collective dynamics for a pair of coupled oscillators subject to a morphogen gradient.
  \item Fig S3: Schematic diagram illustrating the canonical Notch signaling pathway that allows contact-mediated interaction between cells.
  \item Fig S4: Relative importance of \textit{trans} and \textit{cis} forms of Notch-Delta binding for reproducing the observed
sequence of transitions during somitogenesis.
  \item Fig S5: Effect of delay in Notch-Delta mediated interaction on the collective dynamics of a pair of coupled
oscillators responding to an exponential gradient of morphogen concentration cell signaling.
  \item Fig S6: Temporal evolution of the concentration of a morphogen along the gradient as sensed by a particular
cell in the growing PSM.
  \item Fig S7: Temporal evolution of the dynamical variables in a pair of coupled oscillators responding to an
exponential gradient of morphogen concentration.
  \item Fig S8: The collective dynamics of a pair of coupled oscillators separated by a distance $\Delta x$ responding to
  an exponential gradient of morphogen concentration having steepness $\lambda_g$.
  \item Fig S9: Effect of noise on the collective dynamics of a pair of oscillators interacting via Notch-Delta coupling
modulated by an exponential gradient of morphogen concentration.
  \item Fig S10: Collective dynamics of a cellular array represented in terms of the bound Notch concentration ($N^{b}$).
  \item Fig S11: The model dynamics can generate temporally invariant spatial patterns with different periodicities.
  \item Fig S12:  A two-dimensional cellular array can exhibit a linear sequence
  of alternating peaks and troughs of the protein concentration $Y$.
\end{enumerate}

\section*{Modeling the dynamics of clock gene expression in cells coupled via Notch-Delta signaling}
\begin{figure*}[ht]
\includegraphics[width=\textwidth]{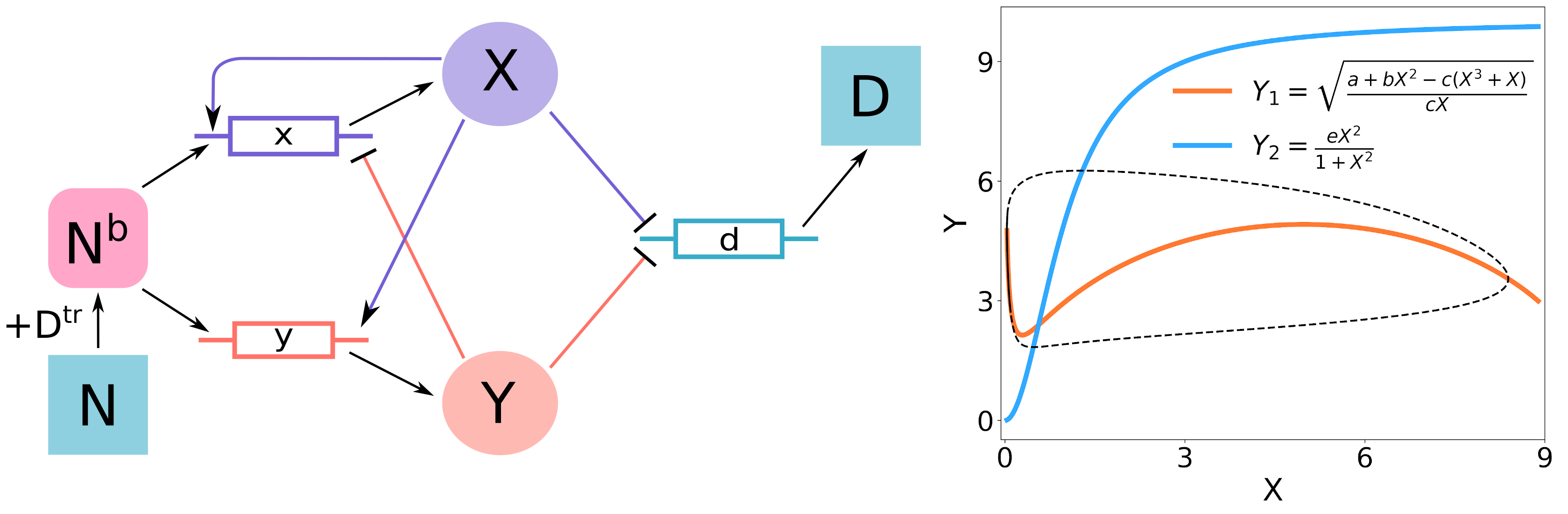}
\caption{{\bf Schematic diagram of the genetic oscillator model and phase portrait of its dynamics.}
[left] Schematic diagram of the interactions resulting in
gene expression oscillations in the model used in the main text.
The system comprises an activator gene $x$ and an inhibitor gene $y$
that yield the gene products $X$ and $Y$, respectively. These in turn
inhibit the expression of Delta gene $d$, resulting in suppression of the
production of Delta ligands $D$. Each cell contains Notch receptors $N$
that, when bound with Delta ligands from another cell, causes the
Notch Intracellular Domain (NICD, represented through the proxy
variable $N^b$) to cleave off and act as
transcriptional factors that upregulate the expression of
the downstream genes $x$ and $y$.
[right] The dynamics of expression levels $X$ and $Y$ for the activator and
inhibitor genes, respectively, in a single uncoupled oscillator, shown
in terms of the trajectory of the limit cycle (broken curve) and the
nullclines (red: $\dot{X}=0$, blue: $\dot{Y}=0$) obtained from Eqs.~(1)-(2)
in the main text.}
\label{figS1}
\end{figure*}

In the model presented in the main text, we consider a two-component genetic oscillator~\cite{guantes2006v2}, comprising an activator gene $x$ and an inhibitor gene $y$. The parameter values of the model (see the next section) have been chosen such that it exhibits autonomous oscillatory activity. The interactions between these two genes are schematically shown in Fig.~\ref{figS1}~(left). Each uncoupled oscillator consists of two variables $X$ and $Y$ that represent the concentrations of the products expressed by genes $x$ and $y$, respectively. The trajectory of an uncoupled oscillator in the $X$-$Y$ phase plane is shown in Fig.~\ref{figS1}~(right), along with the nullclines $\dot{X}=0$ and $\dot{Y}=0$. Fig.~\ref{figS1}~(left) also displays the nature of the interactions between genes $x$ and $y$, where the variables $N^{b}$, $N$ and $D$ describe the Notch-Delta coupling between cells~\cite{sprinzak2010v2}.

\begin{figure*}[ht]
\includegraphics[width=\textwidth]{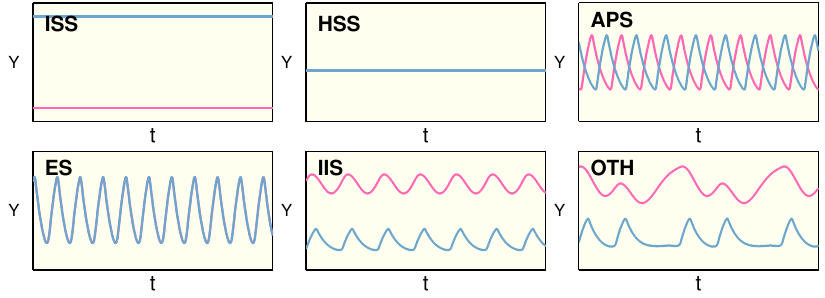}
\caption{{\bf Representative patterns of collective dynamics for a pair of coupled
oscillators subject to a morphogen gradient.}
Typical time series of the inhibitor gene expression $Y$ for
a pair of cells coupled to each other (each exhibiting oscillations
in isolation, as shown in Fig.~\ref{figS1}). Different dynamical regimes
obtained by varying the coupling parameters $f$, $g$ and $h$ that are shown
here correspond to (top row, left) Inhomogeneous
Steady State (ISS), (bottom row, left) Exact Synchronization (ES),
(top row, center) Homogeneous Steady
State (HSS), (bottom row, center) Inhomogeneous In-phase Synchronization (IIS),
(top row, right) Anti-Phase Synchronization (APS) and (bottom row, right)
Other synchronization patterns (OTH). The activity of the two cells
are represented using two different colors in each panel. Note that the
two curves overlap in the ES and HSS regimes.}
\label{figS2}
\end{figure*}

In a system of coupled oscillators, we observe a variety of synchronization behavior (examples of which are shown in Fig.~\ref{figS2}).
These can be categorized into  $6$ principal collective dynamical patterns, namely Inhomogeneous Steady State (ISS), Exact Synchronization (ES), Homogeneous Steady State (HSS), Inhomogeneous In-phase Synchronization (IIS), Anti-Phase Synchronization (APS) and Other synchronization patterns (OTH).

\clearpage

\section*{Model Parameters}
As described in the main text, the time-evolution of each cell $i$ (coupled to its neighbors via
Notch-Delta coupling) is described by the following set of coupled differential
equations:
\begin{figure*}[ht]
  \centering
   \includegraphics[width=0.7\textwidth]{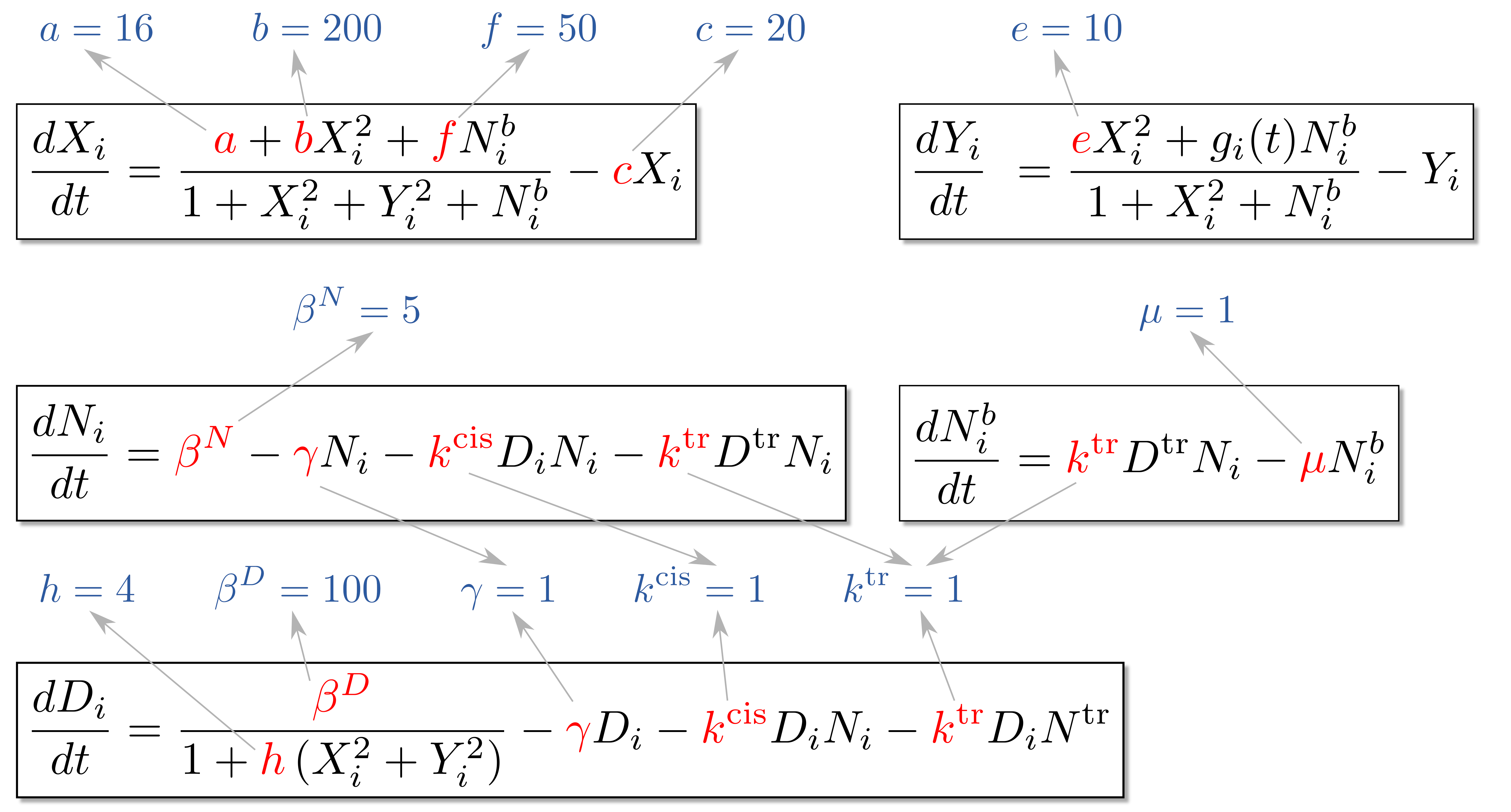}
   \label{parameter_fig}
\end{figure*}

The model parameters, described in the table below, can be broadly classified into those that specify the properties of
(i) the cellular oscillator comprising the clock genes, (ii) Notch signaling between neighboring
cells that couples their respective activities, and (iii) the morphogen gradient that modulates
the strength of the Notch-mediated coupling. The values chosen for each of these
parameters (indicated above the corresponding equations in which they appear)
are selected to ensure that the gene circuit in each cell exhibits oscillations in the absence
of any coupling to neighboring cells and that the time-scale of Notch signaling dynamics is approximately the same as that of the cellular oscillator. The spatial gradient of morphogen concentration has been kept temporally invariant.
Note that the parameter $g_i (t)$ which represents the strength with which Notch promotes
the expression of one of the clock genes ($Y$) varies in space (as a result of the
morphogen gradient), as well as, in time (due to the effective movement of a cell
along the morphogen gradient towards the anterior because of the expansion of the PSM via the addition of new cells at its posterior end).
\begin{table}[H]
\centering
\begin{tabular}{|c|l|}\hline
Parameter & Interpretation \\\hline
$a$ & Base transcription rate of gene $X$ \\
$b$ & Transcription rate of gene $X$ on self-activation (i.e., dimer of protein
expressed by $X$ binds to promoter of gene $X$) \\
$c$ & Degradation rate of the product of gene $X$ \\
$e$ & Transcription rate of gene $Y$ when the dimer of protein expressed by $X$ is bound to the promoter of gene $Y$ \\
$f$ & Transcription rate of gene $X$ when $N^b$ is bound to the promoter of gene $X$ \\
$g$ & Transcription rate of gene $Y$ when $N^b$ is bound to the promoter of gene $Y$ \\
$\beta^N$ & Maximal production rate of the free Notch receptors \\
$\beta^D$ & Maximal production rate of the Delta ligands \\
$k^{cis}$ & Forward rate constant for complex formation involving Notch receptor and Delta ligand of the same cell \\
$k^{tr}$ & Forward rate constant for complex formation between Notch receptor of a cell
and Delta ligand of its neighbor\\
$\gamma$ & Degradation rate of the Notch receptor \\
$\mu$ & Degradation rate of the Notch Intracellular Domain, NICD ($N^b$) \\
$h$ & Affinity of Delta repressing dimers of proteins
expressed by $X$ and $Y$ for the promoter of gene coding Delta ligand\\\hline
\end{tabular}
\label{tableS1}
\end{table}
\clearpage
\section*{Numerical solution of the model}
As the dynamics of each cell is described by a system of five coupled ordinary differential equations
(ODEs), to describe the behavior of a $1$D array of $N$ cells, we solve $5N$ coupled ODEs using
an adaptive numerical integrator (such as, \texttt{odeint} from the Python \texttt{scipy} module).
The initial values of the dynamical variables are chosen from an uniform distribution over the
unit interval $[0,1]$.
\section*{Relative contributions of {\it trans} and {\it cis} binding to the collective dynamics}
\begin{figure*}[ht]
  \centering
  \includegraphics[width=0.7\textwidth]{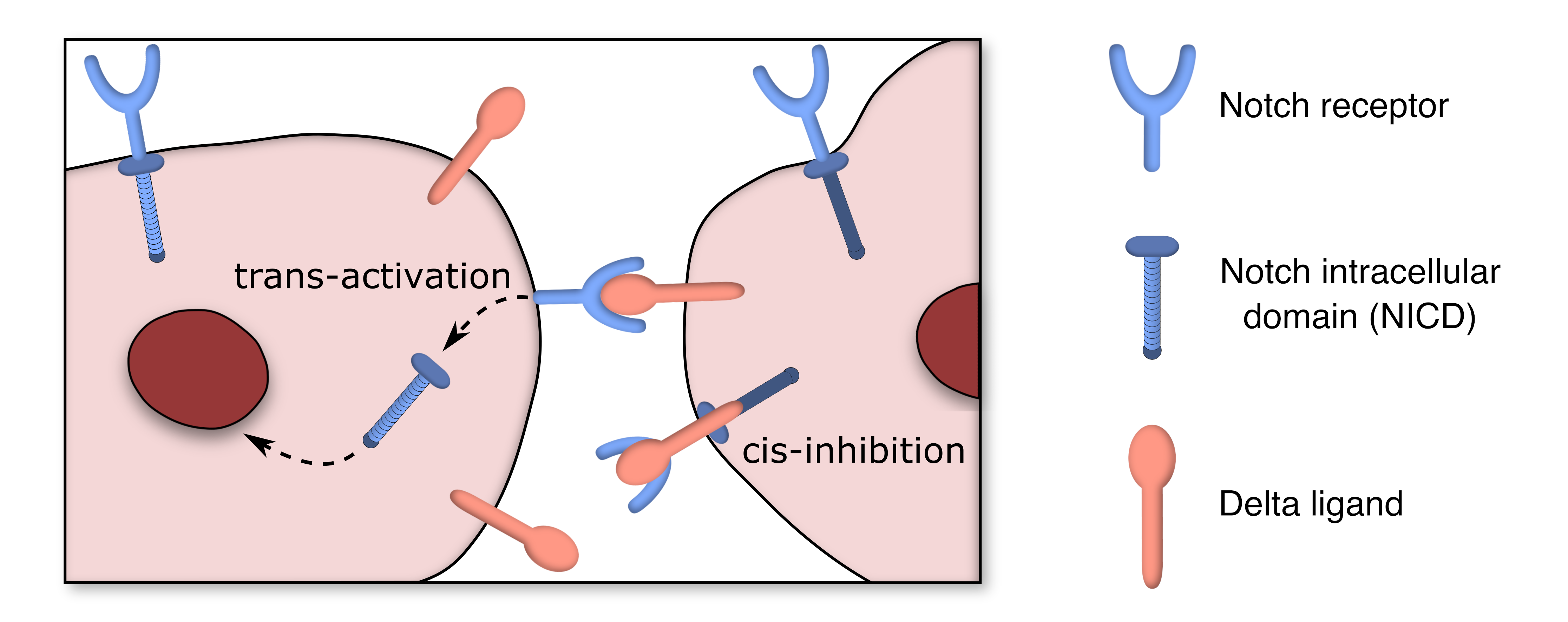}
  \caption{Schematic diagram illustrating the canonical Notch signaling pathway that allows
  contact-mediated interaction between cells.}
  \label{simple_notch}
\end{figure*}

The Notch-Delta signaling pathway (Fig.~\ref{simple_notch}) provides neighboring cells in the presomitic mesoderm (PSM) a mechanism for communicating with each other. Binding of the Notch receptor of a cell to a
Delta ligand belonging to a neighboring cell leads to the cleavage of the Notch intra-cellular
domain (NICD) which serves as a transcription factor for downstream genes either directly or
indirectly. This process is referred to as \textit{trans}-activation of Notch to distinguish
it from \textit{cis}-inhibition in which Delta and Notch of the same cell forms
a complex resulting in the ligand and the receptor no longer being available for inter-cellular
communication.

\begin{figure*}[ht]
  \centering
\includegraphics[width=0.7\textwidth]{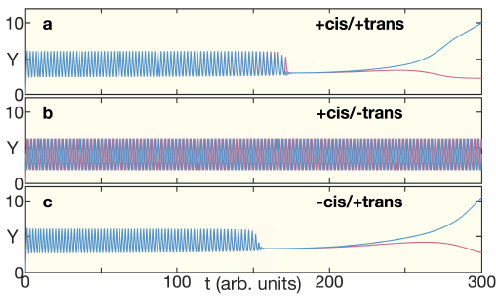}
\caption{{\bf Relative importance of {\it trans} and {\it cis} forms of Notch-Delta binding for reproducing the observed sequence of transitions during somitogenesis.}
The results reported in the main text assume that both
forms of binding are operative with strengths $k^{\rm tr} = k^{\rm cis} = 1$ (a). (b) In the absence of
{\it trans} binding (i.e., $k^{\rm tr} = 0$), synchronized oscillations will continue indefinitely,
indicating that this form of binding is crucial for the cessation of oscillations and subsequent polarization.
(c) In contrast, when the {\it cis} form of the binding is absent (i.e., $k^{\rm cis} = 0$), we observe
no qualitative difference in the collective dynamics compared to the situation when both
types of coupling are present [see (a)].}
\label{trans_cis}
\end{figure*}

Fig.~\ref{trans_cis} demonstrates the relative importance of \textit{trans} and \textit{cis} forms
of Notch-Delta binding in generating the sequence of dynamical transitions observed during somitogenesis. Absence
of \textit{cis}-inhibition leads to only minor, quantitative changes in the cell activity
from that observed when \textit{cis}-
and \textit{trans}-bindings both have the same propensity. However, in the absence of \textit{trans}-activation the cells are no longer able to communicate with each other, preventing the system
from exhibiting any of the transitions associated with somitogenesis.

%
\clearpage

\section*{The effect of delay}
\begin{figure*}[ht]
\includegraphics[width=\textwidth]{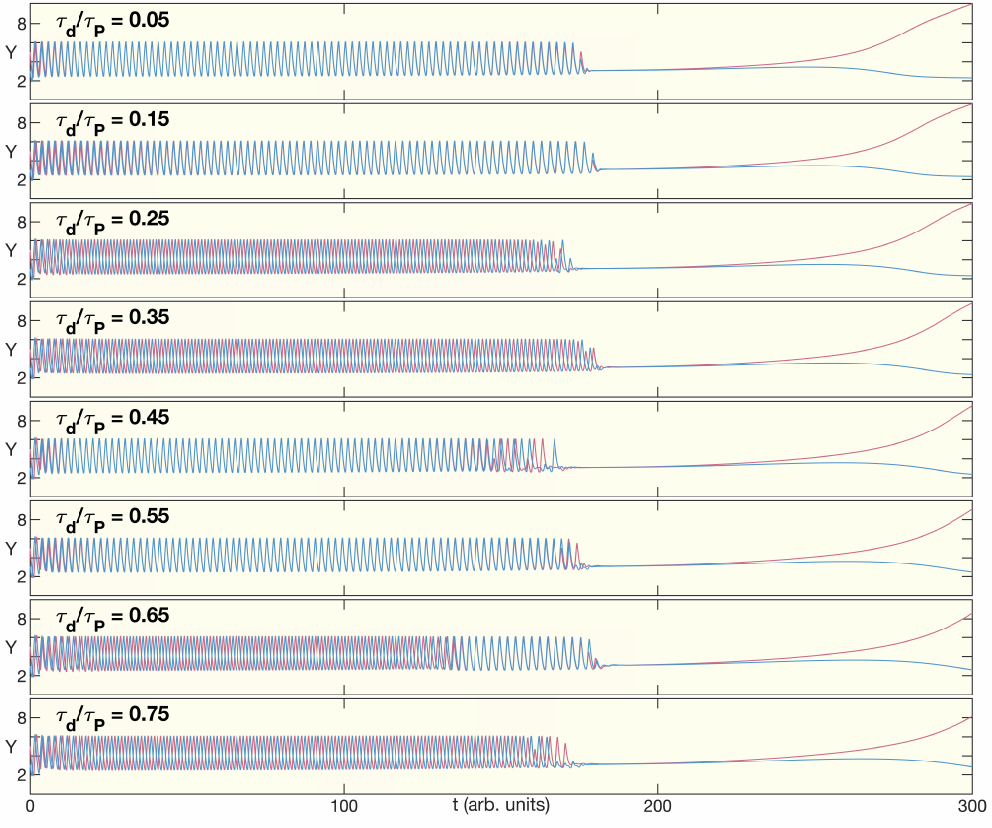}
\caption{{\bf Effect of delay in Notch-Delta mediated interaction on the
collective dynamics of a pair of coupled oscillators responding to
an exponential gradient of morphogen concentration cell signaling.}
The effect of the NICD (represented by the proxy variable $N_b$)
on $X$ and $Y$, as well as the effect of $X$ and $Y$ on the
expression of the ligand Delta $D$, are each delayed by a
period $\tau_{d}$. The results shown here are obtained for a range of
values of $\tau_{d}$, expressed relative to $\tau_{P}$, the period of
an uncoupled oscillator.
We observe that the transition from
oscillations to an inhomogeneous steady state is unaffected by the
delay.}
\label{fig_delay}
\end{figure*}

In order to investigate the effect of an explicit delay in the system, we incorporate a delay of duration $\tau_{d}$ in both (i) the regulation of $X$ and $Y$ by $N^{b}$ and (ii) the repression of the ligand Delta  by $X$ and $Y$. In Fig.~\ref{fig_delay}, we show the effect of increasing delay, which is expressed
in terms of the relative magnitude $\tau_{d}/\tau_{P}$, where $\tau_{P}$ is the period of the uncoupled oscillator, on the dynamics of
the system. We observe the same qualitative nature for the dynamical transitions as seen in Fig.~4~(d) of the main text,
suggesting that our results are robust with respect to incorporation of signaling delays.

\clearpage

\section*{Modeling the morphogen gradient in a growing PSM}
\begin{figure*}[ht]
\includegraphics[width=0.7\textwidth]{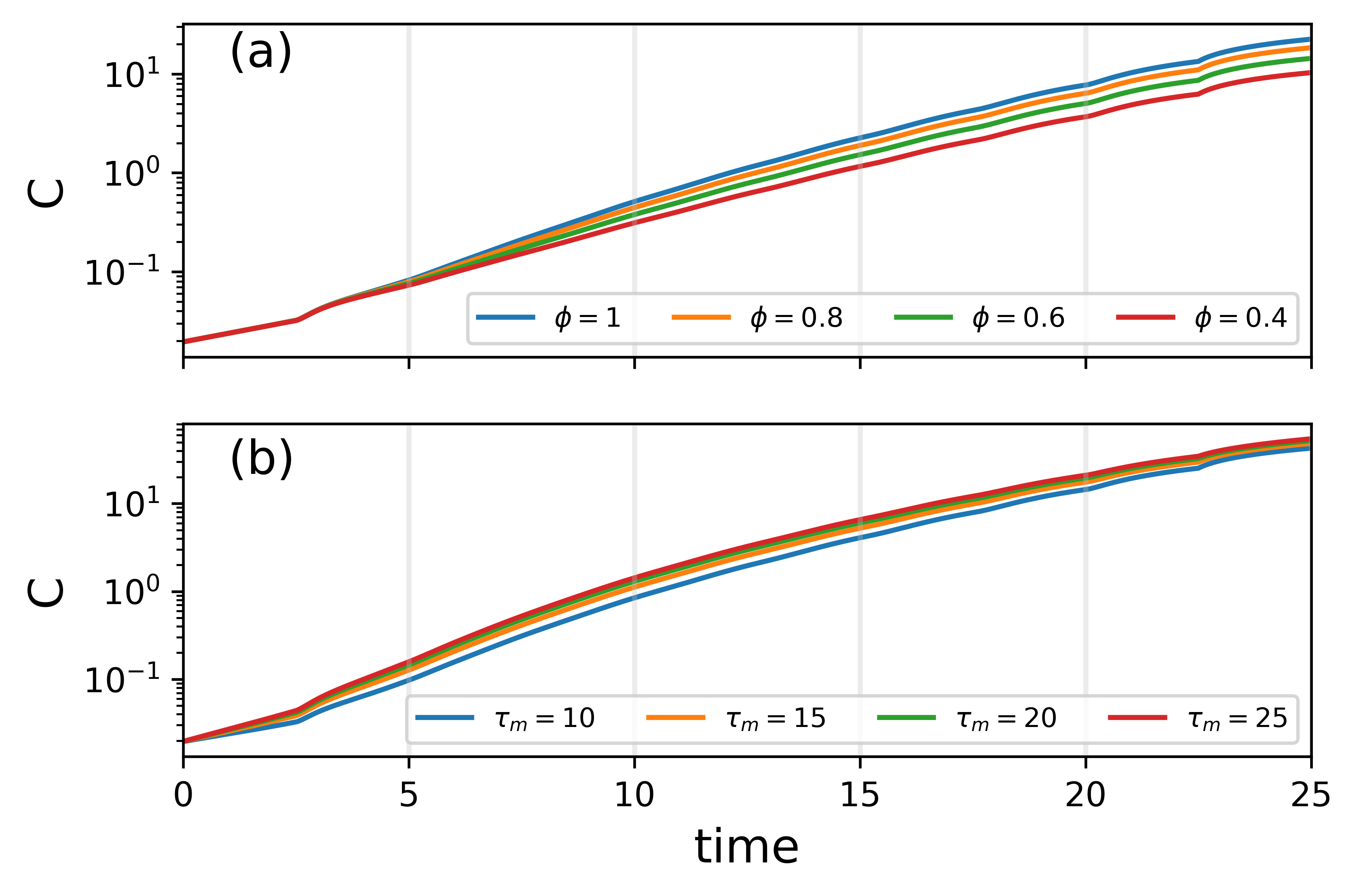}
\caption{{\bf Temporal evolution of the concentration of a morphogen along the gradient as sensed by a particular cell in the growing PSM.}
Concentration of the morphogen gradient as sensed by a particular cell in the growing PSM with time, shown for (a) different
values of $\phi$ for $\tau_m = 10$, and (b) different values of $\tau_m$ for
$\phi = 0.4$. The simulation domain comprises $N=10$ putative somites at any given time, while the other parameter
values are $D=1$, $\tau_d=5$ and $R(t=0) = 1.0$.}
\label{fig_grad}
\end{figure*}

A crucial element of our model is the gradient of morphogen concentration across the presomitic mesoderm (PSM).
We have explicitly considered the morphogen to be Retinoic acid (RA) whose concentration is highest at the anterior end of the PSM.
Experimental  evidence suggests that as the anteriorly located newly formed somites mature, each of them serve as
a source of RA~\cite{del2003v2,del2004v2,rhinn2012v2}.
This suggests that a
source, diffusion and linear degradation (SDD) model for RA will provide an appropriate description for the variation of
the morphogen concentration across space and time. In particular, the source grows in strength
as the number of mature somites increases over time. If one considers the PSM to be growing in discrete
steps, with each new mature somite (which secretes RA at a rate $\phi$) being added after a time interval $\tau_d$,
the concentration $C_i$ of the morphogen at each putative somite in position $i$ on the anteroposterior axis of the growing PSM can be described by the differential
equation:
\begin{equation*}
\frac{d C_i}{d t} = R(t) \delta_{i,0} + D (C_{i+1} + C_{i-1} - 2 C_i ) - \frac{C_i}{\tau_m},
\end{equation*}
where the time-varying production term $R(t)$ increases in a step-like manner by $\phi$ after each interval of duration $\tau_d$,
$\delta_{i,0}$ is a Kronecker delta function indicating that the source is at the anterior end of the domain (using
the simplifying assumption of a point source rather than a distributed one), and
$D$ and $\tau_m$ are the effective diffusion rate and average lifetime of the $RA$ molecules, respectively.
We note that qualitatively identical results are obtained if, instead of increasing in discrete time steps, the production
rate $R(t)$ changes in a continuous fashion, for example, as described by the differential equation
\begin{equation*}
\frac{dR}{dt} = \frac{\phi}{\tau_d}.
\end{equation*}
Fig.~\ref{fig_grad} shows the resulting exponential profile of the morphogen concentration that will be experienced
by a particular cell in the expanding PSM, which validates the use of an exponential profile for the morphogen in the
main text.

\clearpage
\section*{Temporal evolution of all the variables in case of a pair of coupled cells}
\begin{figure*}[ht]
\includegraphics[width=\textwidth]{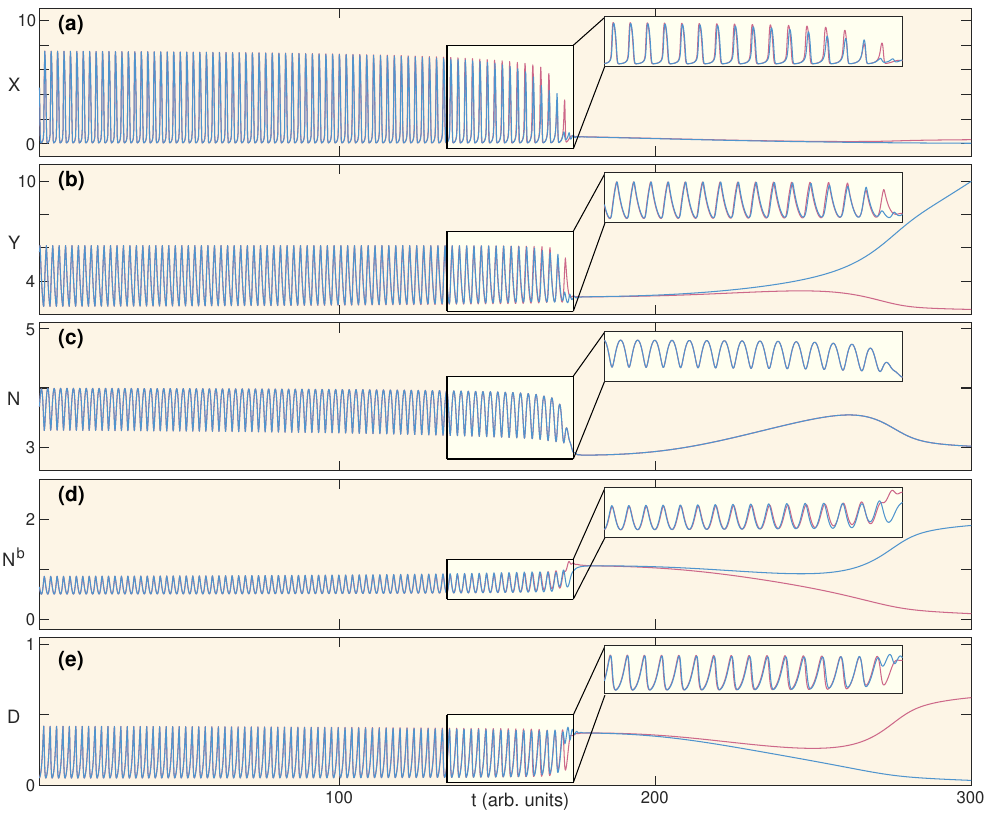}
\caption{{\bf Temporal evolution of the dynamical variables in a
pair of coupled oscillators responding to an exponential gradient of morphogen
concentration.}
Unless mentioned otherwise, the following parameter values have been used for
all model simulations: $a = 16.0$, $b = 200.0$, $c = 20.0$, $e = 10.0$, $\beta^N = 5.0$, $\beta^D = 100.0$, $\gamma =1.0$, $k^{tr} = 1.0$, $k^{cis} = 1.0$, $\mu = 1.0$, $f = 50.0$, $h = 4$, $g_{in} = 0.5$, $g_{max} = 15.0$, $dx = 3.0 $ and $t_{max} = 300$.}
\label{figallvar}
\end{figure*}

In the main text, the time-evolution of only one of the variables (namely, the expression of
the clock gene $Y$) has been shown as representative of the dynamics of the system.
As can be seen from Fig.~\ref{figallvar}, all the other variables also exhibit qualitatively
identical behavior, with the exception of free Notch $N$ whose production is
not dependent on any of the other dynamical variables describing the state of the system.
\clearpage

\section*{The effect of steepness of the morphogen gradient}
\begin{figure*}[ht]
\includegraphics[width=0.7\textwidth]{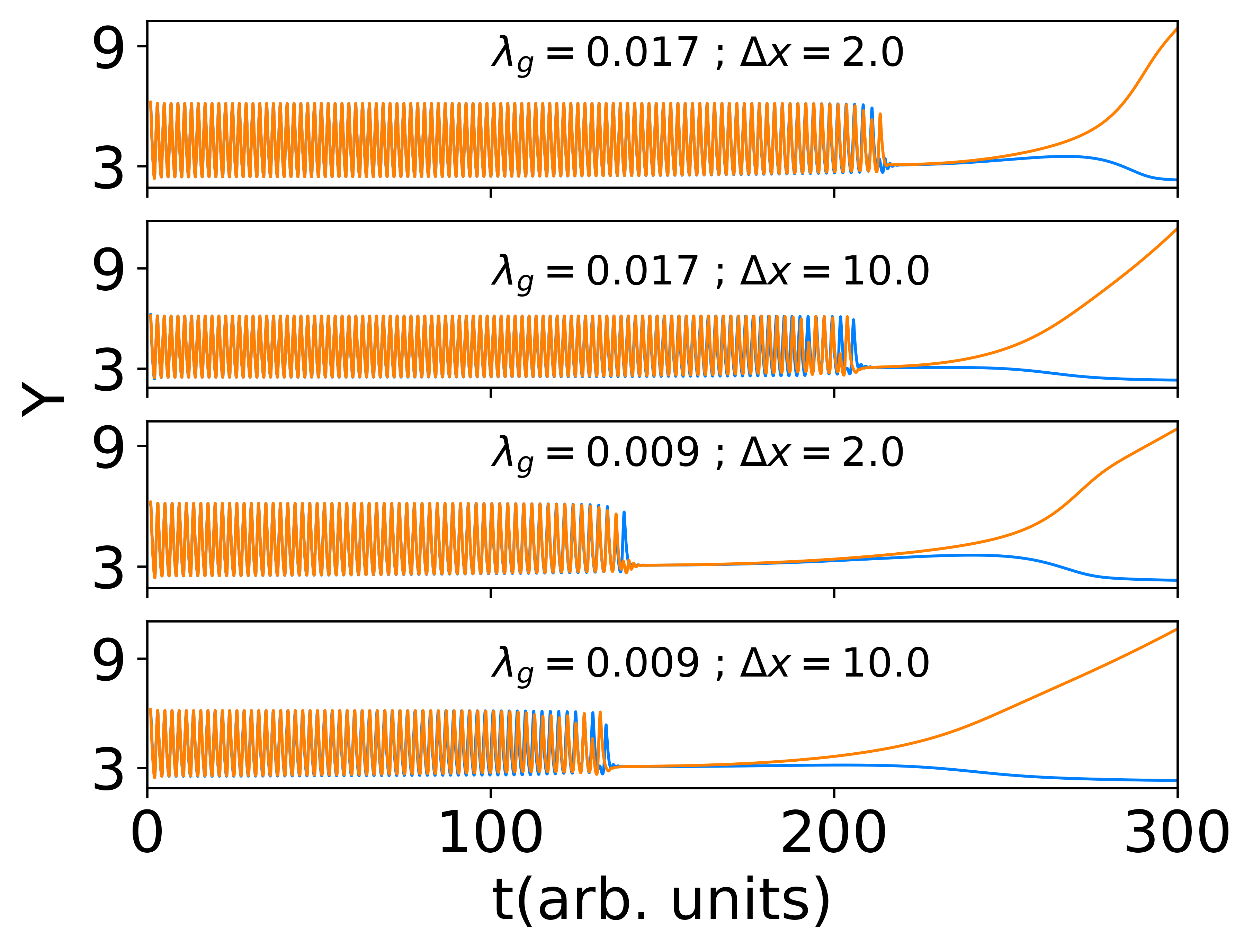}
\caption{{\bf The collective dynamics of a pair of coupled oscillators separated by a distance $\Delta x$ responding to
an exponential gradient of morphogen concentration having steepness $\lambda_g$.}
In comparison
to shallower gradients (e.g., $\lambda_g = 0.009$), steeper gradients (e.g., $\lambda_g = 0.017$)
prolong the oscillatory phase and correspondingly reduce the duration of HSS prior to the cells attaining
different fates characterized by high and low values of protein concentration.
A larger spatial separation between the centers of the two cells results in an earlier initiation of ISS, further reducing the HSS duration.
Note that the qualitative features of the sequence of transitions from ES to ISS shown in Fig.~4(d) in the main text remain unchanged.}
\label{fig_steep}
\end{figure*}

As shown in Fig.~\ref{fig_grad}, the morphogen concentration increases exponentially from the posterior to the anterior along
the AP axis of the PSM, such that the concentration at the position $x$, $g(x) \sim \exp (\lambda_g x)$. The exponent $\lambda_g$
can be tuned to vary the steepness of the morphogen gradient. In Fig.~\ref{fig_steep} we consider a coupled pair of cellular oscillators
separated by a distance $\Delta x$ (see the schematic in Fig.~4~(a) of the main text) which are moving along the gradient.
While the sequence of transitions are qualitatively identical to those shown in the main text,
we observe that higher values of $\lambda_g$, which increases the slope of the gradient, results in the duration of the oscillatory regime
being prolonged. More importantly, it substantially reduces the time the system spends in the transient homogeneous state
prior to the ISS phase corresponding to onset of polarization. A higher value of $\Delta x$ implies a larger spatial interval over
which the concentration changes, which effectively alters the slope of the morphogen gradient sensed by the cells. Increasing $\Delta x$
results in the ISS phase being initiated even earlier, thereby making it appear that the transition from oscillation arrest to onset of polarization
is immediate.

\clearpage

\section*{The effect of noise}
\begin{figure*}[ht]
\includegraphics[width=0.7\textwidth]{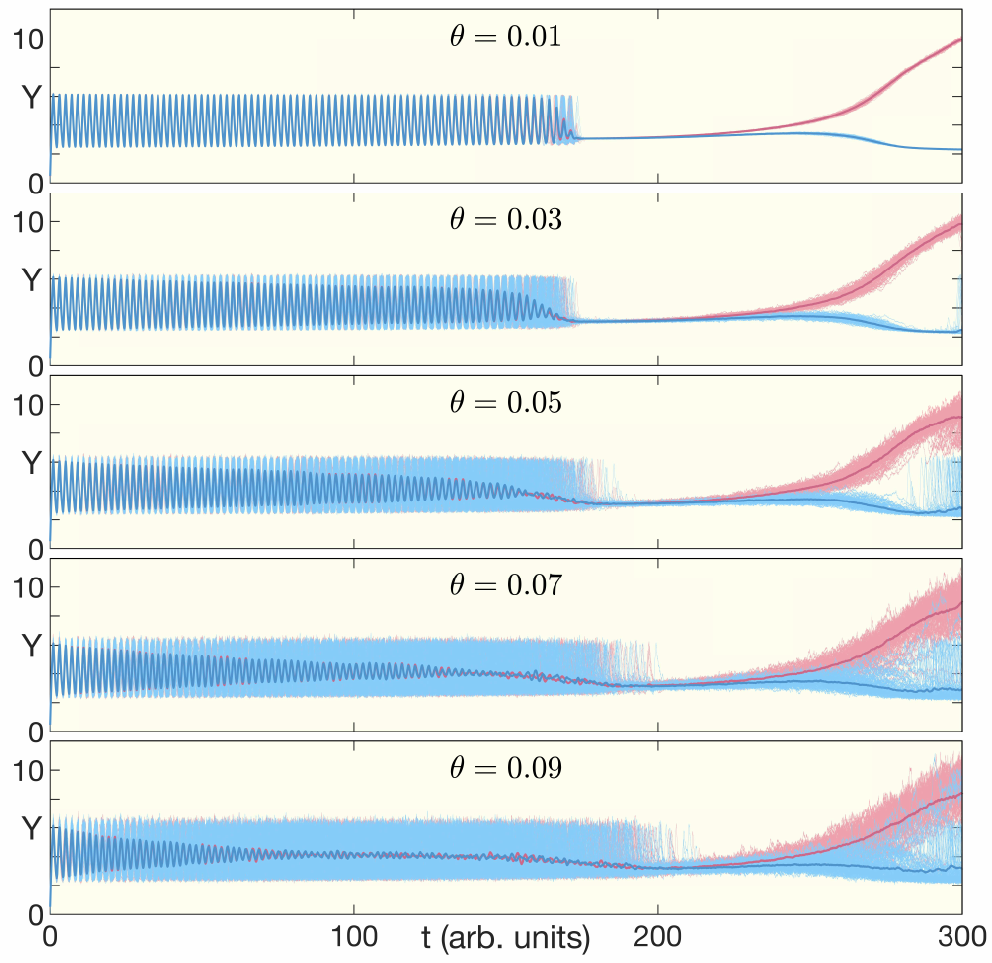}
\caption{{\bf Effect of noise on the collective dynamics of a pair of oscillators interacting via Notch-Delta coupling
modulated by an exponential gradient of morphogen concentration.}
The evolution equation of each of
the five variables $V \in \{X,Y,N,N^b,D\}$ is augmented with a noise whose variance is proportional to the instantaneous value
of the corresponding variable. This is implemented by introducing an additional term which is the
product of Gaussian noise $\mathcal{N} (0,1)$ with the variable $V$ and a parameter $\theta$ which
quantifies the strength of the noise. The other parameter values used in the evolution equations are identical to that
for the deterministic simulation shown in Fig.~4(d) in the main text. Each panel shows the results
of $100$ realizations of stochastic simulations for a given $\theta$. The averages
of these values for each cell calculated over the different realizations are the trajectories
indicated using the thicker curves. We note that, although increasing noise results in higher
dispersion around the mean trajectories, the qualitative features of the sequence of transitions
from ES to ISS shown in Fig.~3(d) remain unchanged.}
\label{fignoise}
\end{figure*}

As gene expression in inherently noisy, it is important to verify that the sequence of transitions shown by the model introduced here are
robust with respect to fluctuations in the concentrations of the relevant variables. We have, therefore, investigated the role of noise by
incorporating stochasticity in the evolution equations for each of the variables $X,Y,N,N^b,D$. This is done by adding a term which
is the product of the variable with unit Gaussian noise and a parameter $\theta$ which quantifies the strength of the noise.
Fig.~\ref{fignoise} shows that while increasing $\theta$ results in larger deviations around the mean trajectories (indicated by the
thicker curves), the qualitative features of the sequence of transitions seen using the deterministic model remain invariant.

\clearpage

\section*{Temporal evolution of bound Notch concentration}
\begin{figure*}[ht]
\includegraphics[width=0.9\textwidth]{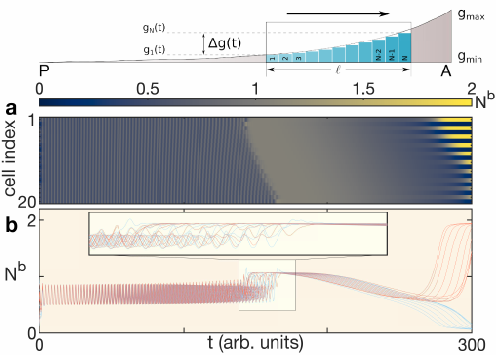}
\caption{{\bf Collective dynamics of a cellular array represented in terms of the bound Notch concentration ($N^{b}$).}
The collective dynamics of a cellular array comprising $20$ cells responding
to an exponential gradient of morphogen concentration
reproduces the spatio-temporal evolution of PSM activity seen
during somitogenesis, similar to Fig.~5 in the main text, but showing
bound Notch ($N^{b}$) instead of the inhibitor variable $Y$.}
\label{figNb}
\end{figure*}

In the main text, the behavior of the system has been characterized by the time-evolution
of expression of $Y$, one of the clock genes. Fig.~\ref{figNb} shows that a qualitatively
similar picture is observed if instead one considers the time-evolution of another variable,
namely, the concentration of NICD ($N^b$) released as a result of \textit{trans}-activation of Notch
receptors.
\clearpage
\section*{The model dynamics can generate temporally invariant spatial patterns with different periodicities}
\begin{figure*}[ht]
\includegraphics[width=\textwidth]{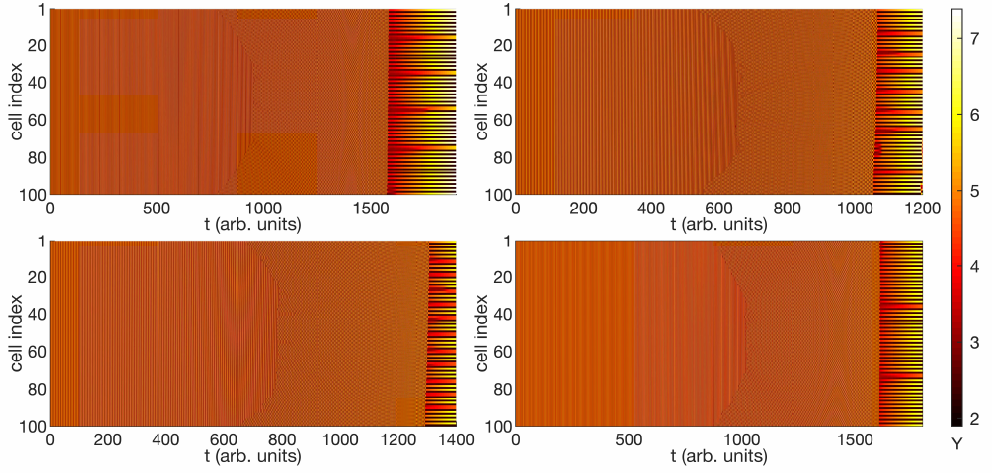}
\caption{{\bf The model dynamics can generate temporally invariant
spatial patterns with different periodicities.}
ISS states that deviate from those shown in the main text are obtained
by considering faster signaling dynamics, specifically, increasing the values of
$\beta^N$, $\beta^D$, $k^{tr}$, $k^{cis}$, $\gamma$ and $\mu$ by a factor of 10 from those
used for the results reported in the main text. The different panels correspond to distinct sets of
values used for the parameters $f$, $h$, $g_{max}$, $dx$ and $t_{max}$.
The patterns are shown in a 1D segment comprising $N=100$ cells.}
\label{fig_band}
\end{figure*}

The results reported in the main text exhibit patterns that correspond to somites, each of
which comprise
two cells. However, the model dynamics is capable of a richer set of possible patterns
having different periodicities that can be obtained by using different values for the model
parameters. This can be seen from Fig.~\ref{fig_band} which shows a representative
sample of the different types of spatio-temporal evolution leading to patterns with distinct
periodicities.
\clearpage
\section*{A two-dimensional cellular array can exhibit a linear sequence of alternating peaks and troughs of
the protein concentration}
\begin{figure*}[ht]
\includegraphics[width=0.7\textwidth]{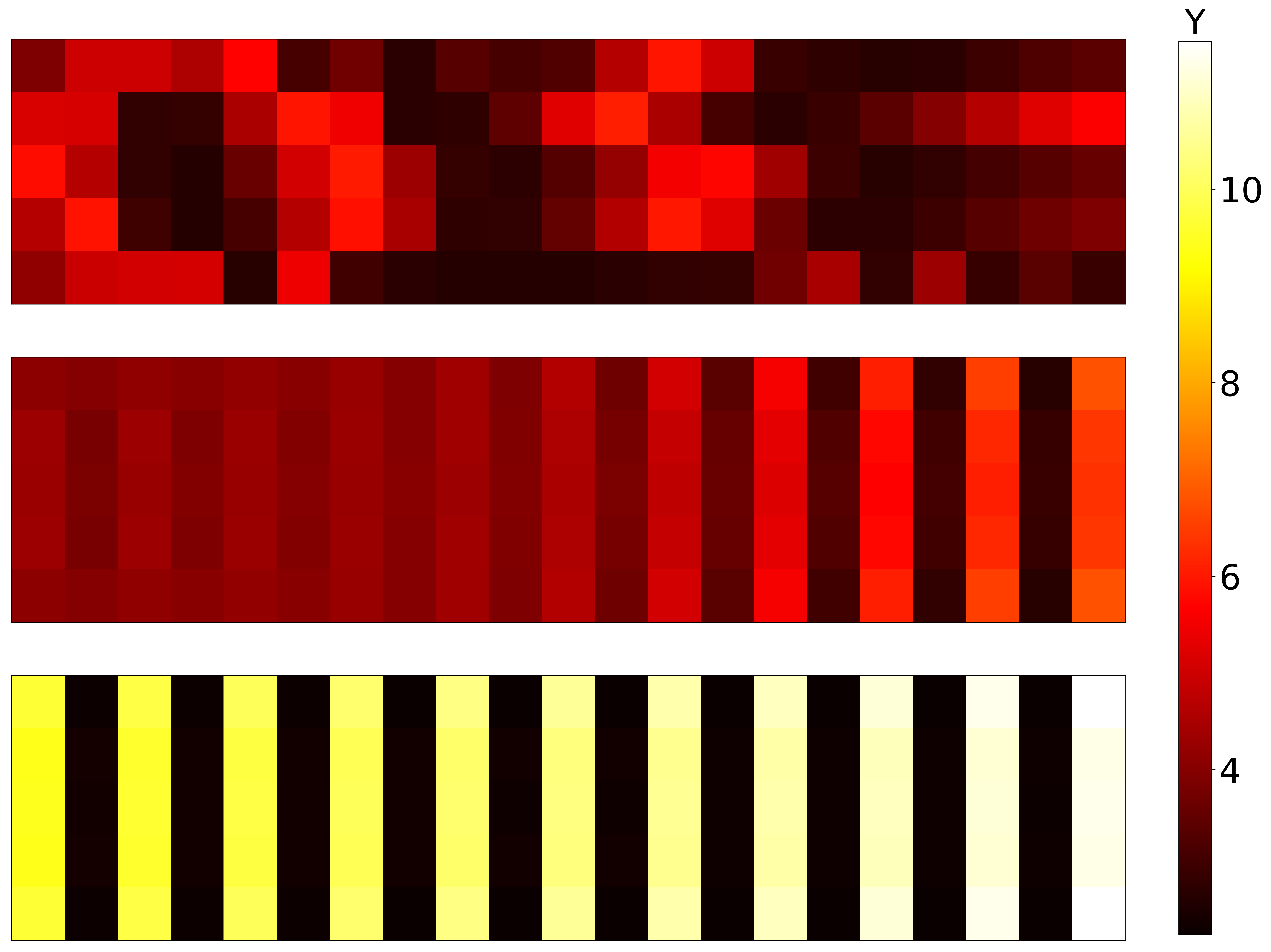}
\caption{{\bf A two-dimensional cellular array can exhibit a linear sequence
of alternating peaks and troughs of the protein concentration $Y$.}
Anisotropic inter-cellular communication, which can result from the area of contact between
neighboring cells being larger along the AP axis compared to that along its transverse,
yields an ISS state where columns of cells aligned perpendicular to the AP axis exhibit the
same concentration of $Y$, thereby attaining the same fate. In contrast, neighboring cells
along the AP axis attain different fates characterized by high and low values of $Y$. The
panels above show the concentration $Y$ across the domain at different points of time,
viz., (top) $t = 350$, (center) $t=850$ and (bottom) $t=1000$ arb. units}
\label{figtwoD}
\end{figure*}

In the main text we have reported only the behavior of a linear array of cells.
However, we have explicitly verified that a two-dimensional array of cells
which are coupled anisotropically can give rise to patterns that resemble the
$1$D patterns (Fig.~\ref{figtwoD}). The anisotropy is a consequence of the cellular
geometry, with the relatively smaller area of contact along the axis perpendicular to
the AP axis resulting in a lower intensity of contact-mediated signal between cells
across the transverse direction.


\newpage
\begin{thebibliography}{99}

\bibitem{Koch1994}
A.~J. Koch and H.~Meinhardt.
\newblock Biological pattern formation: from basic mechanisms to complex
  structures.
\newblock {\em Rev Mod Phys}, 66:1481--1507, 1994.

\bibitem{dequeant2008}
M.-L. Dequ{\'e}ant and O. Pourqui{\'e}.
\newblock Segmental patterning of the vertebrate embryonic axis.
\newblock {\em Nat Rev Genet}, 9(5):370, 2008.

\bibitem{gomez2008}
C. Gomez, E.~M.~{\"O}zbudak, J. Wunderlich, D.
  Baumann, J. Lewis, and O. Pourqui{\'e}.
\newblock Control of segment number in vertebrate embryos.
\newblock {\em Nature}, 454(7202):335, 2008.

\bibitem{vonk2008}
F.~J.~Vonk and M.~K.~Richardson.
\newblock Developmental biology: Serpent clocks tick faster.
\newblock {\em Nature}, 454(7202):282, 2008.

\bibitem{Gilbert2013}
S.~F.~Gilbert.
\newblock {\em Developmental Biology}.
\newblock Sinauer, 2013.

\bibitem{stollewerk2003}
A. Stollewerk, M. Schoppmeier, and W.~G.~M.~Damen.
\newblock Involvement of notch and delta genes in spider segmentation.
\newblock {\em Nature}, 423(6942):863, 2003.

\bibitem{sarrazin2012}
A.~F. Sarrazin, A.~D. Peel, and M.~Averof.
\newblock A segmentation clock with two-segment periodicity in insects.
\newblock {\em Science}, 336:338--341, 2012.

\bibitem{pourquie2001}
O. Pourqui{\'e} and P.~P.~L.~Tam.
\newblock A nomenclature for prospective somites and phases of cyclic gene
  expression in the presomitic mesoderm.
\newblock {\em Dev Cell}, 1(5):619--620, 2001.

\bibitem{cooke1976}
J. Cooke and E.~C.~Zeeman.
\newblock A clock and wavefront model for control of the number of repeated
  structures during animal morphogenesis.
\newblock {\em J Theor Biol}, 58(2):455--476, 1976.

\bibitem{pourquie2003}
O. Pourqui{\'e}.
\newblock The segmentation clock: converting embryonic time into spatial
  pattern.
\newblock {\em Science}, 301(5631):328--330, 2003.

\bibitem{baker2006}
R.~E.~Baker, S.~Schnell, and P.~K.~Maini.
\newblock A clock and wavefront mechanism for somite formation.
\newblock {\em Dev Biol}, 293(1):116--126, 2006.

\bibitem{santillan2008}
M. Santill{\'a}n and M.~C.~Mackey.
\newblock A proposed mechanism for the interaction of the segmentation clock
  and the determination front in somitogenesis.
\newblock {\em PLOS ONE}, 3(2):e1561, 2008.

\bibitem{nagahara2009}
H. Nagahara, Y. Ma, Y. Takenaka, R. Kageyama, and K. Yoshikawa.
\newblock Spatiotemporal pattern in somitogenesis: A non-turing scenario with
  wave propagation.
\newblock {\em Phys Rev E}, 80(2):021906, 2009.

\bibitem{hester2011}
S.~D.~Hester, J.~M.~Belmonte, J.~S.~Gens, S.~G.~Clendenon, and J.~A.~Glazier.
\newblock A multi-cell, multi-scale model of vertebrate segmentation and somite
  formation.
\newblock {\em PLOS Comput Biol}, 7(10):e1002155, 2011.

\bibitem{ares2012}
S. Ares, L.~G.~Morelli, D.~J.~J{\"o}rg, A.~C.~Oates, and F. J{\"u}licher.
\newblock Collective modes of coupled phase oscillators with delayed coupling.
\newblock {\em Phys Rev Lett}, 108(20):204101, 2012.

\bibitem{murray2013}
P.~J.~Murray, P.~K.~Maini, and R.~E.~Baker.
\newblock Modelling delta-notch perturbations during zebrafish somitogenesis.
\newblock {\em Dev Biol}, 373:407--421, 2013.

\bibitem{jorg2014}
D.~J.~J{\"o}rg, L.~G.~Morelli, S. Ares, and F. J{\"u}licher.
\newblock Synchronization dynamics in the presence of coupling delays and phase
  shifts.
\newblock {\em Phys Rev Lett}, 112:174101, 2014.

\bibitem{wiedermann2015}
G. Wiedermann, R.~A.~Bone, J.~C.~Silva, M. Bjorklund,
  P.~J.~Murray, and J.~K.~Dale.
\newblock A balance of positive and negative regulators determines the pace of
  the segmentation clock.
\newblock {\em eLife}, 4:e05842, 2015.

\bibitem{Oates2012}
A.~C.~Oates, L.~G.~Morelli, and S. Ares.
\newblock Patterning embryos with oscillations: structure, function and
  dynamics of the vertebrate segmentation clock.
\newblock {\em Development}, 139(4):625--639, 2012.

\bibitem{palmeirim1997}
I. Palmeirim, D. Henrique, D. Ish-Horowicz, and O. Pourqui{\'e}.
\newblock Avian hairy gene expression identifies a molecular clock linked to
  vertebrate segmentation and somitogenesis.
\newblock {\em Cell}, 91(5):639--648, 1997.

\bibitem{dale2000}
K.~J.~Dale and O. Pourqui{\'{e}}.
\newblock A clock-work somite.
\newblock {\em Bioessays}, 22(1):72--83, 2000.

\bibitem{saga2001}
Y. Saga and H. Takeda.
\newblock The making of the somite: molecular events in vertebrate
  segmentation.
\newblock {\em Nat Rev Genet}, 2(11):835, 2001.

\bibitem{maroto2003}
M. Maroto, J.~K.~Dale, M.-L. Dequeant, A.-C. Petit, and O. Pourqui{\'{e}}.
\newblock Synchronised cycling gene oscillations in presomitic mesoderm cells
  require cell-cell contact.
\newblock {\em Int J Dev Biol}, 49(2-3):309--315, 2003.

\bibitem{masamizu2006}
Y. Masamizu, T. Ohtsuka, Y. Takashima, H. Nagahara,
  Y. Takenaka, K. Yoshikawa, H. Okamura, and R. Kageyama.
\newblock Real-time imaging of the somite segmentation clock: revelation of
  unstable oscillators in the individual presomitic mesoderm cells.
\newblock {\em Proc Natl Acad Sci USA}, 103(5):1313--1318, 2006.

\bibitem{riedel2007}
I.~H.~Riedel-Kruse, C.~M{\"u}ller, and A.~C.~Oates.
\newblock Synchrony dynamics during initiation, failure, and rescue of the
  segmentation clock.
\newblock {\em Science}, 317(5846):1911--1915, 2007.

\bibitem{schroter2012}
C. Schr{\"o}ter, S. Ares, L.~G.~Morelli, A. Isakova, K.
  Hens, D. Soroldoni, M. Gajewski, F. J{\"u}licher, S.~J.~Maerkl, B.~Deplancke, and
  A.~C.~Oates.
\newblock Topology and dynamics of the zebrafish segmentation clock core
  circuit.
\newblock {\em PLOS Biol}, 10(7):e1001364, 2012.

\bibitem{webb2016}
A.~B Webb, I.~M.~Lengyel, D.~J.~J{\"o}rg, G. Valentin, F.
  J{\"u}licher, L.~G.~Morelli, and A.~C.~Oates.
\newblock Persistence, period and precision of autonomous cellular oscillators
  from the zebrafish segmentation clock.
\newblock {\em eLife}, 5:e08438, 2016.

\bibitem{jiang1998}
Y.-J. Jiang, L. Smithers, and J. Lewis.
\newblock Vertebrate segmentation: the clock is linked to notch signalling.
\newblock {\em Curr Biol}, 8(24):R868--R871, 1998.

\bibitem{ferjentsik2009}
Z. Ferjentsik, S. Hayashi, J.~K.~Dale, Y. Bessho, A.~Herreman,
  B. De~Strooper, G. del Monte, J.~L. de~la Pompa, and M.
  Maroto.
\newblock Notch is a critical component of the mouse somitogenesis oscillator
  and is essential for the formation of the somites.
\newblock {\em PLoS Genet}, 5(9):e1000662, 2009.

\bibitem{hubaud2014}
A. Hubaud and O. Pourqui\'{e}.
\newblock Signalling dynamics in vertebrate segmentation.
\newblock {\em Nat Rev Mol Cell Bio}, 15:709--721, 2014.

\bibitem{conlon1995}
R.~A.~Conlon, A.~G.~Reaume, and J. Rossant.
\newblock Notch1 is required for the coordinate segmentation of somites.
\newblock {\em Development}, 121(5):1533--1545, 1995.

\bibitem{pourquie1999}
O. Pourqui{\'{e}}.
\newblock Notch around the clock.
\newblock {\em Curr Opin Genet Dev}, 9(5):559--565, 1999.

\bibitem{yun2000}
Y.-J. Jiang, B.~L.~Aerne, L. Smithers, C. Haddon, D. Ish-Horowicz, and J. Lewis.
\newblock Notch signalling and the synchronization of the somite segmentation
  clock.
\newblock {\em Nature}, 408(6811):475, 2000.

\bibitem{lai2004}
E.~C. Lai.
\newblock Notch signaling: control of cell communication and cell fate.
\newblock {\em Development}, 131:965--973, 2004.

\bibitem{huppert2005}
S.~S.~Huppert, M.~X.~G.~Ilagan, B. De~Strooper, and R. Kopan.
\newblock Analysis of notch function in presomitic mesoderm suggests a
  $\gamma$-secretase-independent role for presenilins in somite
  differentiation.
\newblock {\em Dev Cell}, 8(5):677--688, 2005.

\bibitem{mara2007}
A. Mara and S.~A.~Holley.
\newblock Oscillators and the emergence of tissue organization during zebrafish
  somitogenesis.
\newblock {\em Trends Cell Biol}, 17(12):593--599, 2007.

\bibitem{kageyama2007}
R. Kageyama, Y. Masamizu, and Y. Niwa.
\newblock Oscillator mechanism of notch pathway in the segmentation clock.
\newblock {\em Dev Dynam}, 236(6):1403--1409, 2007.

\bibitem{sprinzak2010}
D. Sprinzak, A. Lakhanpal, L. LeBon, L.~A.~Santat, M.~E.~Fontes,
  G.~A.~Anderson, J. Garcia-Ojalvo, and M.~B.~Elowitz.
\newblock Cis-interactions between notch and delta generate mutually exclusive
  signalling states.
\newblock {\em Nature}, 465(7294):86, 2010.

\bibitem{sprinzak2011}
D. Sprinzak, A. Lakhanpal, L. LeBon, J. Garcia-Ojalvo, and
  M.~B.~Elowitz.
\newblock Mutual inactivation of notch receptors and ligands facilitates
  developmental patterning.
\newblock {\em PLOS Comput Biol}, 7(6):e1002069, 2011.

\bibitem{lewis2003}
J. Lewis.
\newblock Autoinhibition with transcriptional delay: a simple mechanism for the
  zebrafish somitogenesis oscillator.
\newblock {\em Curr Biol}, 13(16):1398--1408, 2003.

\bibitem{giudicelli2004}
F. Giudicelli and J. Lewis.
\newblock The vertebrate segmentation clock.
\newblock {\em Curr Opin Genet Dev}, 14:0--414, 2004.

\bibitem{horikawa2006}
K. Horikawa, K. Ishimatsu, E. Yoshimoto, S. Kondo, and H.
  Takeda.
\newblock Noise-resistant and synchronized oscillation of the segmentation
  clock.
\newblock {\em Nature}, 441(7094):719, 2006.

\bibitem{Giudicelli2007}
F. Giudicelli, E.~M.~{\"O}zbudak, G.~J.~Wright, and
  J. Lewis.
\newblock {Setting the tempo in development: an investigation of the zebrafish
  somite clock mechanism}.
\newblock {\em PLOS Biol}, 5(6):e150, 2007.

\bibitem{tiedemann2012}
H.~B.~Tiedemann, E. Schneltzer, S. Zeiser, B. Hoesel, J.
  Beckers, G.~K.~H.~Przemeck, and M.~H.~de~Angelis.
\newblock From dynamic expression patterns to boundary formation in the
  presomitic mesoderm.
\newblock {\em PLOS Comput Biol}, 8(6):e1002586, 2012.

\bibitem{tiedemann2014}
H.~B.~Tiedemann, E. Schneltzer, S. Zeiser, W. Wurst, J.
  Beckers, G.~K.~H.~Przemeck, M.~H.~de Angelis, and D.
  Thieffry.
\newblock Fast synchronization of ultradian oscillators controlled by
  delta-notch signaling with cis-inhibition.
\newblock {\em PLOS Comput Biol}, 10:e1003843, 2014.

\bibitem{mcgrew1998}
M.~J.~McGrew and O. Pourqui{\'e}.
\newblock Somitogenesis: segmenting a vertebrate.
\newblock {\em Curr Opin Genet Dev}, 8(4):487--493, 1998.

\bibitem{Oginuma2010}
M. Oginuma, Y.~Takahashi, S. Kitajima, M. Kiso, J. Kanno,
  A. Kimura, and Y. Saga.
\newblock {The oscillation of Notch activation, but not its boundary, is
  required for somite border formation and rostral-caudal patterning within a
  somite}.
\newblock {\em Development}, 137(9):1515--1522, 2010.

\bibitem{gibb2010}
S. Gibb, M. Maroto, and J.~K.~Dale.
\newblock The segmentation clock mechanism moves up a notch.
\newblock {\em Trends Cell Biol}, 20(10):593--600, 2010.

\bibitem{dubrulle2001}
J. Dubrulle, M.~J.~McGrew, and O. Pourqui{\'e}.
\newblock Fgf signaling controls somite boundary position and regulates
  segmentation clock control of spatiotemporal hox gene activation.
\newblock {\em Cell}, 106(2):219--232, 2001.

\bibitem{dubrulle2004}
J. Dubrulle and O. Pourquié.
\newblock fgf8 mrna decay establishes a gradient that couples axial elongation
  to patterning in the vertebrate embryo.
\newblock {\em Nature}, 427:419--422, 2004.

\bibitem{dubrulle2004development}
J.~Dubrulle and O.~Pourqui\'{e}.
\newblock Coupling segmentation to axis formation.
\newblock {\em Development}, 131:5783--5793, 2004.

\bibitem{vermot2005}
J. Vermot and O. Pourqui{\'e}.
\newblock Retinoic acid coordinates somitogenesis and left--right patterning in
  vertebrate embryos.
\newblock {\em Nature}, 435(7039):215, 2005.

\bibitem{aulehla2004}
A.~Aulehla.
\newblock Segmentation in vertebrates: clock and gradient finally joined.
\newblock {\em Gene Dev}, 18:2060--2067, 2004.

\bibitem{aulehla2008}
A. Aulehla, W. Wiegraebe, V. Baubet, M.~B.~Wahl, C.
  Deng, M. Taketo, M. Lewandoski, and O. Pourqui{\'e}.
\newblock A $\beta$-catenin gradient links the clock and wavefront systems in
  mouse embryo segmentation.
\newblock {\em Nat Cell Biol}, 10(2):186, 2008.

\bibitem{bajard2014}
L.~Bajard, L.~G. Morelli, S.~Ares, J.~Pecreaux, F.~Julicher, and A.~C. Oates.
\newblock Wnt-regulated dynamics of positional information in zebrafish
  somitogenesis.
\newblock {\em Development}, 141:1381--1391, 2014.

\bibitem{goldbeter2007}
A. Goldbeter, D. Gonze, and O. Pourqui{\'e}.
\newblock Sharp developmental thresholds defined through bistability by
  antagonistic gradients of retinoic acid and fgf signaling.
\newblock {\em Dev Dynam}, 236(6):1495--1508, 2007.

\bibitem{goldbeter2008}
A. Goldbeter and O. Pourqui{\'e}.
\newblock Modeling the segmentation clock as a network of coupled oscillations
  in the notch, wnt and fgf signaling pathways.
\newblock {\em J Theor Biol}, 252(3):574--585, 2008.

\bibitem{mazzitello2008}
K.~I.~Mazzitello, C.~M.~Arizmendi, and H.~G.~E.~Hentschel.
\newblock Converting genetic network oscillations into somite spatial patterns.
\newblock {\em Phys Rev E}, 78(2):021906, 2008.

\bibitem{jorg2016}
D.~J.~J{\"o}rg, A.~C.~Oates, and F. J{\"u}licher.
\newblock Sequential pattern formation governed by signaling gradients.
\newblock {\em Phys Biol}, 13(5):05LT03, 2016.

\bibitem{Youn1980}
B.~W.~Youn, R.~E.~Keller, and G.~M.~Malacinski.
\newblock An atlas of notochord and somite morphogenesis in several anuran and
  urodelean amphibians.
\newblock {\em Development}, 59(1):223--247, 1980.

\bibitem{Tlili2019}
S.~Tlili, J.~Yin, J.-F. Rupprecht, M.~A.~Mendieta-Serrano, G. Weissbart,
  N.~Verma, X.~Teng, Y.~Toyama, J.~Prost, and T.~E.~Saunders.
\newblock Shaping the zebrafish myotome by intertissue friction and active
  stress.
\newblock {\em Proc Natl Acad Sci USA}, 116(51):25430--25439, 2019.

\bibitem{Lander2011}
A.~D.~Lander.
\newblock Pattern, growth, and control.
\newblock {\em Cell}, 144(6):955--969, 2011.

\bibitem{murray2011}
P.~J.~Murray, P.~K.~Maini, and R.~E.~Baker.
\newblock The clock and wavefront model revisited.
\newblock {\em J Theor Biol}, 283(1):227--238, 2011.

\bibitem{Tomka2018}
T. Tomka, D. Iber, and M. Boareto.
\newblock {Travelling waves in somitogenesis: Collective cellular properties
  emerge from time-delayed juxtacrine oscillation coupling}.
\newblock {\em Prog Biophys Mol Biol}, 137:76--87, 2018.

\bibitem{Adhyapok2021}
P. Adhyapok, A. M. Piatkowska M. J. Norman, S. G. Clendenon, C. D. Stern, J. A. Glazier, and J. M. Belmonte.
\newblock A mechanical model of early somite segmentation.
\newblock {\em iScience}, 24(4):102317, 2021

\bibitem{Narayanan2021}
R. Narayanan, M. A. Mendieta-Serrano, and  T. E. Saunders.
\newblock The role of cellular active stresses in shaping the zebrafish body axis.
\newblock {\em Curr Opin Cell Biol}, 73 : 69-77, 2021

\bibitem{Meinhardt1982}
H. Meinhardt.
\newblock {\em {Models of Biological Pattern Formation}}.
\newblock Academic Press, London, 1982.

\bibitem{Francois2007}
P. Fran{\c{c}}ois, V. Hakim, and E.~D.~Siggia.
\newblock {Deriving structure from evolution: metazoan segmentation}.
\newblock {\em Mol Syst Biol}, 3(1):154, 2007.

\bibitem{cotterell2015}
J. Cotterell, A. Robert-Moreno, and J. Sharpe.
\newblock A local, self-organizing reaction-diffusion model can explain somite
  patterning in embryos.
\newblock {\em Cell Syst}, 1:257--269, 2015.

\bibitem{Wahi2016}
K. Wahi, M.~S.~Bochter, and S.~E.~Cole.
\newblock The many roles of notch signaling during vertebrate somitogenesis.
\newblock {\em Semin Cell Dev Biol}, 49:68--75, 2016.

\bibitem{liao2017}
B.-K. Liao and A.~C.~Oates.
\newblock Delta-notch signalling in segmentation.
\newblock {\em Arthropod Struct Dev}, 46(3):429--447, 2017.

\bibitem{Guantes2006}
R. Guantes and J.~F.~Poyatos.
\newblock Dynamical principles of two-component genetic oscillators.
\newblock {\em PLOS Comput Biol}, 2:e30, 2006.

\bibitem{Goodenough2009}
D.~A.~Goodenough and D.~L.~Paul.
\newblock Gap junctions.
\newblock {\em Cold Spring Harb Perspect Biol}, 1(1):a002576, 2009

\bibitem{Takke1999}
C. Takke and J.~A.~Campos-Ortega.
\newblock her1, a zebrafish pair-rule like gene, acts downstream of notch
  signalling to control somite development.
\newblock {\em Development}, 126(13):3005--3014, 1999.

\bibitem{Oates2002}
A.~C.~Oates and R.~K.~Ho.
\newblock Hairy/e (spl)-related (her) genes are central components of the
  segmentation oscillator and display redundancy with the delta/notch signaling
  pathway in the formation of anterior segmental boundaries in the zebrafish.
\newblock {\em Development}, 129(12):2929--2946, 2002.

\bibitem{maroto2005}
M. Maroto, J.~K.~Dale, M.-L. Dequeant, A.-C. Petit, and O.
  Pourquie.
\newblock Synchronised cycling gene oscillations in presomitic mesoderm cells
  require cell-cell contact.
\newblock {\em Int J Dev Biol}, 49:309--315, 2005.

\bibitem{gurdon2001}
J.~B.~Gurdon and P.-Y.~Bourillot.
\newblock Morphogen gradient interpretation.
\newblock {\em Nature}, 413(6858):797, 2001.

\bibitem{aulehla2010}
A. Aulehla and O. Pourqui{\'e}.
\newblock Signaling gradients during paraxial mesoderm development.
\newblock {\em Cold Spring Harb Perspect Biol}, 2(2):a000869, 2010.

\bibitem{sawada2001}
A. Sawada, M. Shinya, Y.-J. Jiang, A. Kawakami, A.
  Kuroiwa, and H. Takeda.
\newblock Fgf/mapk signalling is a crucial positional cue in somite boundary
  formation.
\newblock {\em Development}, 128(23):4873--4880, 2001.

\bibitem{moreno2004}
T.~A.~Moreno and C. Kintner.
\newblock Regulation of segmental patterning by retinoic acid signaling during
  xenopus somitogenesis.
\newblock {\em Dev Cell}, 6(2):205--218, 2004.

\bibitem{Lander2002}
A.~D.~Lander, Q. Nie, and F.~Y.~M.~Wan.
\newblock Do morphogen gradients arise by diffusion?
\newblock {\em Dev Cell}, 2(6):785--796, 2002.

\bibitem{Bergmann2007}
S. Bergmann, O. Sandler, H. Sberro, S. Shnider, E. Schejter, B.-Z.
  Shilo, and N. Barkai.
\newblock Pre-steady-state decoding of the bicoid morphogen gradient.
\newblock {\em PLOS Biol}, 5(2):1--11, 2007.

\bibitem{Barkai2009}
N. Barkai and B.-Z.~Shilo.
\newblock Robust generation and decoding of morphogen gradients.
\newblock {\em Cold Spring Harb Perspect Biol}, 1(5), 2009.

\bibitem{Hubaud2017}
A. Hubaud, I. Regev, L. Mahadevan, and O. Pourqui\'{e}.
\newblock Excitable dynamics and Yap-dependent mechanical cues drive the segmentation clock.
\newblock {\em Cell}, 171(3):668-682, 2017.

\bibitem{delaune2012}
E.~A.~Delaune, P. Fran{\c{c}}ois, N.~P.~Shih, and S.~L.~Amacher.
\newblock {Single-cell-resolution imaging of the impact of Notch signaling and
  mitosis on segmentation clock dynamics}.
\newblock {\em Dev Cell}, 23(5):995--1005, 2012.

\bibitem{shih2015}
N.~P.~Shih, P. Fran{\c{c}}ois, E.~A.~Delaune, and S.~L.~Amacher.
\newblock Dynamics of the slowing segmentation clock reveal alternating
  two-segment periodicity.
\newblock {\em Development}, 142(10):1785--1793, 2015.

\bibitem{del2003}
R.~D.~del Corral, I. Olivera-Martinez, A. Goriely, E. Gale,
  M. Maden, and K. Storey.
\newblock Opposing fgf and retinoid pathways control ventral neural pattern,
  neuronal differentiation, and segmentation during body axis extension.
\newblock {\em Neuron}, 40(1):65--79, 2003.

\bibitem{del2004}
R.~D.~del Corral and K.~G.~Storey.
\newblock Opposing fgf and retinoid pathways: a signalling switch that controls
  differentiation and patterning onset in the extending vertebrate body axis.
\newblock {\em Bioessays}, 26(8):857--869, 2004.

\bibitem{rhinn2012}
M. Rhinn and P. Doll{\'e}.
\newblock Retinoic acid signalling during development.
\newblock {\em Development}, 139(5):843--858, 2012.

\bibitem{Schroter2008}
C. Schr{\"o}ter, L. Herrgen, A. Cardona, G.~J.~Brouhard, B.
  Feldman, and A.~C.~Oates.
\newblock Dynamics of zebrafish somitogenesis.
\newblock {\em Dev Dynam}, 237(3):545--553, 2008.

\bibitem{Hadjivasiliou2016}
Z. Hadjivasiliou, G.~L.~Hunter, and B. Baum.
\newblock A new mechanism for spatial pattern formation via lateral and
  protrusion-mediated lateral signalling.
\newblock {\em J Roy Soc Interface}, 13(124):20160484, 2016.

\bibitem{clark2019}
E. Clark, A.~D.~Peel, and M. Akam.
\newblock Arthropod segmentation.
\newblock {\em Development}, 146(18):dev170480, 2019.

\bibitem{herrgen2010}
L. Herrgen, S. Ares, L.~G.~Morelli, C. Schr{\"o}ter, F.
  J{\"u}licher, and A.~C.~Oates.
\newblock Intercellular coupling regulates the period of the segmentation
  clock.
\newblock {\em Curr Biol}, 20(14):1244--1253, 2010.

\bibitem{Stern2015}
C.~D.~Stern and A.~M.~Piatkowska
\newblock Multiple roles of timing in somite formation.
\newblock {\em Semin Cell Dev Biol}, 42:134-139, 2015.

\end{thebibliography}

\begin{thebibliography}{99}

\bibitem{Guantes2006v2}
R. Guantes and J.~F.~Poyatos.
\newblock Dynamical principles of two-component genetic oscillators.
\newblock {\em PLOS Comput Biol}, 2:e30, 2006.

\bibitem{sprinzak2010v2}
D. Sprinzak, A. Lakhanpal, L. LeBon, L.~A.~Santat, M.~E.~Fontes,
  G.~A.~Anderson, J. Garcia-Ojalvo, and M.~B.~Elowitz.
\newblock Cis-interactions between notch and delta generate mutually exclusive
  signalling states.
\newblock {\em Nature}, 465(7294):86, 2010.

\bibitem{del2003v2}
R.~D.~del Corral, I. Olivera-Martinez, A. Goriely, E. Gale,
  M. Maden, and K. Storey.
\newblock Opposing fgf and retinoid pathways control ventral neural pattern,
  neuronal differentiation, and segmentation during body axis extension.
\newblock {\em Neuron}, 40(1):65--79, 2003.

\bibitem{del2004v2}
R.~D.~del Corral and K.~G.~Storey.
\newblock Opposing fgf and retinoid pathways: a signalling switch that controls
  differentiation and patterning onset in the extending vertebrate body axis.
\newblock {\em Bioessays}, 26(8):857--869, 2004.

\bibitem{rhinn2012v2}
M. Rhinn and P. Doll{\'e}.
\newblock Retinoic acid signalling during development.
\newblock {\em Development}, 139(5):843--858, 2012.

\end{thebibliography}
\end{document}